\documentclass{article}
\usepackage{arxiv}

\usepackage{microtype}
\usepackage{amsmath,amsfonts}
\usepackage{algorithm}
\usepackage{array}
\usepackage{textcomp}
\usepackage{stfloats}
\usepackage{url}
\usepackage{verbatim}
\usepackage{graphicx}
\usepackage{subfigure}
\usepackage{cite}
\usepackage[colorlinks=true,linkcolor=blue,citecolor=blue,urlcolor=blue]{hyperref}
\usepackage{cleveref}
\usepackage{ulem}
\usepackage{orcidlink}
\usepackage{xcolor}
\usepackage{multirow}
\usepackage{booktabs}
\usepackage{caption}
\usepackage{tabularx}
\usepackage{mathptmx} 
\usepackage{tabu}
\usepackage{booktabs}   
\usepackage{mwe}
\usepackage{algpseudocode}

\algnewcommand\algorithmicswitch{\textbf{switch}}
\algnewcommand\algorithmiccase{\textbf{case}}
\algnewcommand\algorithmicassert{\texttt{assert}}
\algnewcommand\Assert[1]{\State \algorithmicassert(#1)}%
\algdef{SE}[SWITCH]{Switch}{EndSwitch}[1]{\algorithmicswitch\ #1\ \algorithmicdo}{\algorithmicend\ \algorithmicswitch}%
\algdef{SE}[CASE]{Case}{EndCase}[1]{\algorithmiccase\ #1}{\algorithmicend\ \algorithmiccase}%
\algtext*{EndSwitch}%
\algtext*{EndCase}%
\newcommand{\methodname}{VOICE}
\newcommand{\appname}{VOICE}
\usepackage{xcolor}
\definecolor{DRonecolor}{HTML}{EA9999}
\definecolor{DRtwocolor}{HTML}{F9CB9C}
\definecolor{DRthreecolor}{HTML}{A2C4C9}
\definecolor{DRfourcolor}{HTML}{B6D7A8}
\definecolor{DRfivecolor}{HTML}{B4A7D6}
\definecolor{DRsixcolor}{HTML}{DD7E6B}
\definecolor{DRsevencolor}{HTML}{D5A6BD}

\newcommand{\DRone}{\textbf{\textcolor{DRonecolor}{DR1--Steering}}}

\newcommand{\DRtwo}{\textbf{\textcolor{DRtwocolor}{DR2--View}}}

\newcommand{\DRthree}{\textbf{\textcolor{DRthreecolor}{DR3--Context}}}

\newcommand{\DRfour}{\textbf{\textcolor{DRfourcolor}{DR4--Style}}}

\newcommand{\DRfive}{\textbf{\textcolor{DRfivecolor}{DR5--Guiding}}}

\newcommand{\DRsix}{\textbf{\textcolor{DRsixcolor}{DR6--Adaptability}}}

\newcommand{\DRseven}{\textbf{\textcolor{DRsevencolor}{DR7--Flexibility}}}

\title{VOICE: Visual Oracle for Interaction, Conversation, and Explanation}

\author{ \href{https://orcid.org/0000-0002-1358-8718}{\includegraphics[scale=0.06]{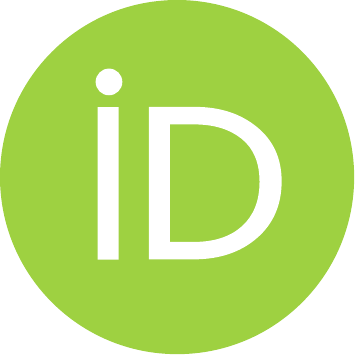}\hspace{1mm}Donggang Jia$^{1,*}$},\href{https://orcid.org/0009-0004-1021-8887}{\includegraphics[scale=0.06]{orcid.pdf}\hspace{1mm}Alexandra Irger$^{1,2,*}$},
\href{https://orcid.org/0000-0002-7207-1276}{\includegraphics[scale=0.06]{orcid.pdf}\hspace{1mm}Lonni Besançon$^{3}$},
\href{https://orcid.org/0000-0002-8077-4692}{\includegraphics[scale=0.06]{orcid.pdf}\hspace{1mm}Ond\v{r}ej Strnad$^{1}$},\And
\href{https://orcid.org/0000-0003-4610-8730}{\includegraphics[scale=0.06]{orcid.pdf}\hspace{1mm}Deng Luo$^{1}$},
\href{https://orcid.org/0000-0001-8503-0118}{\includegraphics[scale=0.06]{orcid.pdf} 
\hspace{1mm}Johanna Björklund$^{4}$},
\href{https://orcid.org/0000-0002-9466-9826}{\includegraphics[scale=0.06]{orcid.pdf}\hspace{1mm}Anders Ynnerman$^{3}$}, and \href{https://orcid.org/0000-0003-4248-6574}{\includegraphics[scale=0.06]{orcid.pdf}\hspace{1mm}Ivan Viola$^{1}$}\\
$^{1}$King Abdullah University of Science and Technology (KAUST), Saudi Arabia.\\
E-mail: \{donggang.jia\,$|$\,ondrej.strnad\,$|$\,deng.luo\,$|$\,ivan.viola\}@kaust.edu.sa. \\
$^{2}$TU Wien, Austria.\\
E-Mail: alexandra.irger@tuwien.ac.at. \\
$^{3}$Linköping University, Sweden. \\
E-mail: \{lonni.besancon\,$|$\,anders.ynnerman\}@liu.se.\\
$^{4}$Umeå University, Sweden.\\
E-mail: johanna.bjorklund@umu.se.\\
$^{*}$D. Jia and A. Irger are co-first authors.
}

\renewcommand{\shorttitle}{VOICE}

\hypersetup{
pdftitle={VOICE: Visual Oracle for Interaction, Conversation, and Explanation},
pdfsubject={q-bio.NC, q-bio.QM},
pdfauthor={Donggang Jia, Alexandra Irger, Ond\v{r}ej Strnad, Johanna Björklund, Anders Ynnerman, and Ivan Viola},
pdfkeywords={Conversational visualization, multi-scale data, explanatory visualization},
}

\begin{document}
\maketitle

\begin{abstract}
We present VOICE, a novel approach to science communication that connects large language models' (LLM) conversational capabilities with interactive exploratory visualization. VOICE introduces several innovative technical contributions that drive our conversational visualization framework. Our foundation is a pack-of-bots that can perform specific tasks, such as assigning tasks, extracting instructions, and generating coherent content. We employ fine-tuning and prompt engineering techniques to tailor bots' performance to their specific roles and accurately respond to user queries. Our interactive text-to-visualization method generates a flythrough sequence matching the content explanation. Besides, natural language interaction provides capabilities to navigate and manipulate the 3D models in real-time. The VOICE framework can receive arbitrary voice commands from the user and respond verbally, tightly coupled with corresponding visual representation with low latency and high accuracy. We demonstrate the effectiveness of our approach by applying it to the molecular visualization domain: analyzing three 3D molecular models with multi-scale and multi-instance attributes. We finally evaluate VOICE with the identified educational experts to show the potential of our approach. All supplemental materials are available at \url{https://osf.io/g7fbr}. 
\end{abstract}


\section{Introduction}
\begin{figure}[t]
    \centering
    \includegraphics[width=\linewidth]{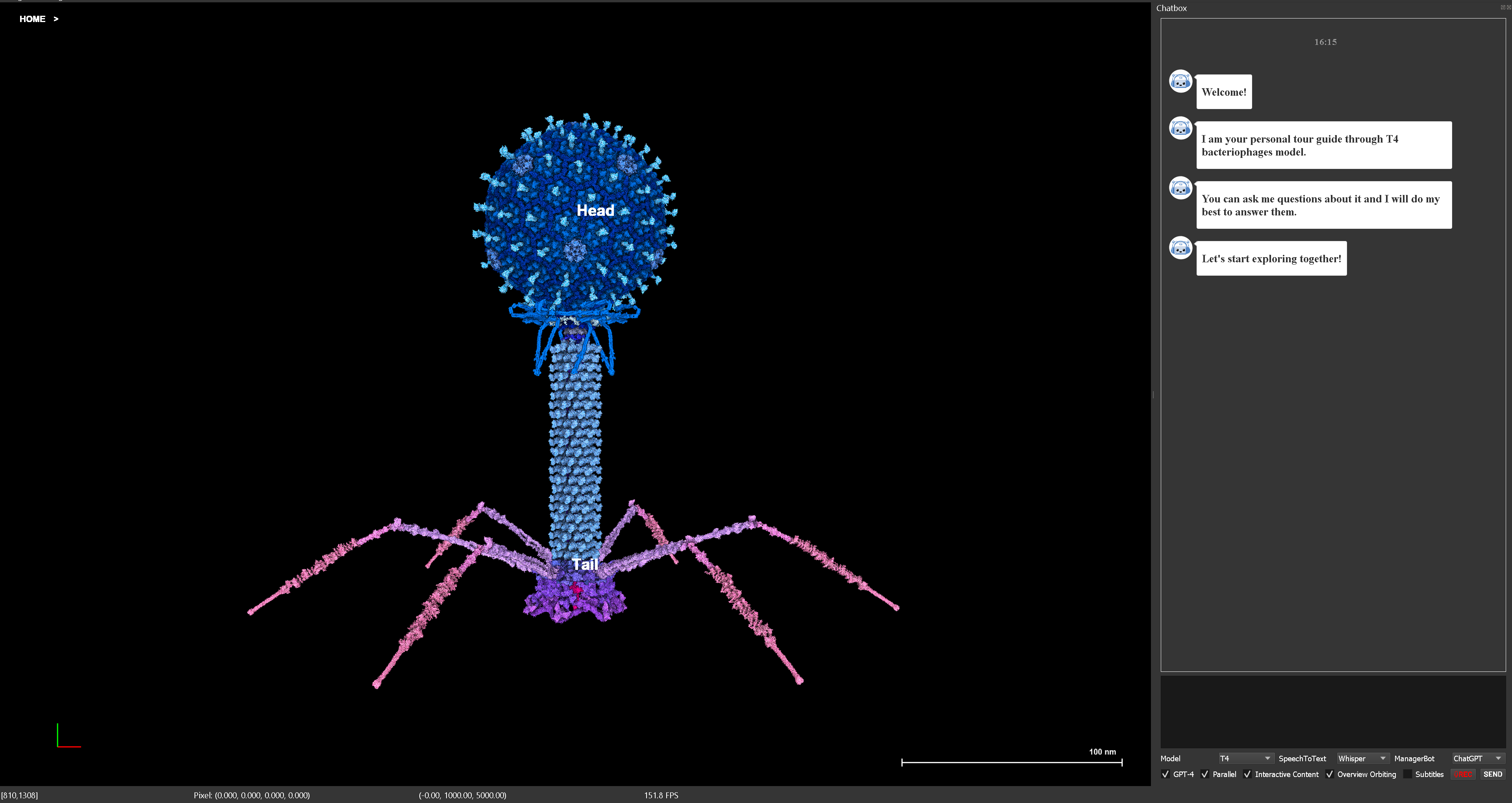}
    \caption{\textbf{VOICE's initial screen.} 
    VOICE can process an arbitrary speech request to answer a question, return a corresponding animation, or conversationally explore the model.
    }
    \label{fig:teaser}
\end{figure}

Science communication is undergoing rapid changes that are driven by greater availability of data,  increasing capabilities of commodity computational resources, and, most importantly, methodological advances in research areas such as visualization and interaction. Many researchers devote themselves to bringing scientific findings from the research frontier to the public by, for example, visualizing the raw data generated in biological or physical experiments. A major difficulty is that non-expert audiences typically lack the knowledge needed to interpret the visual representations without external guidance. This has motivated the introduction of interactive visualization in exhibits in public venues such as science centers. A confluence of explanatory and exploratory visualization has followed~\cite{ynnerman:18:exploranation} and manifested in several installations with high public impact~\cite{ynnerman:16:inside}. 

The complexity and non-linearity of learning processes are well documented~\cite{scanlon2011technology} and not to be underestimated. Excessive exploratory freedom can lead to engagement thresholds for science center visitors that are hard to overcome. Several mitigating strategies have been tried, including intuitive interaction interfaces, simplified visual representations, embedded instructions, and storytelling~\cite{Ynnerman2020}. It is clear that visual representations and linguistic descriptions are indispensable in the learning process. It is widely regarded~\cite{Krange2020PTG} that only knowledgeable guides can assist visitors in using the full potential of interactive visualization and create conversational experiences that adapt to visitors’ needs. However, the cost of human resources needed to meet all visitors at a large-scale science center vastly exceeds the typical staffing budget. Even with unlimited funding,  the scarcity of skilled facilitators remains a bottleneck. Audiences, therefore, would greatly benefit from interactive tools for explanation and demonstration that provide enough freedom to complement human guides. 

Recent advances in generative language modeling~\cite{openai_chatgpt_2022} have opened new possibilities for AI-human conversations by leveraging the vast sources of knowledge available online. In this paper, we present a step towards tightly coupled interactive visual and verbal conversations in public spaces. Our knowledge domain is in molecular biology, which exhibits complex multi-scale structures and processes requiring in-depth explanation and exploration~\cite{kouril2021molecumentary}. We prompt language models to process user instructions and answer biology questions to ensure user interaction freedom. Our method can process arbitrary instructions and integrate a real-time visual display with accurate textual domain knowledge. The visual representation method benefits from the hierarchical molecule structure and can navigate through a 3D molecule model according to the user's high-level instructions.

In this work, we collaborate closely with educational experts from Norrköping's Visualization Center, who have been working in the space of public dissemination and informal learning contexts, to develop VOICE, a novel conversational visualization tool. We conduct a formative study with the aforementioned experts to identify the design requirements and validate our system prototype. The feedback is overwhelmingly positive and frames VOICE as an exciting new resource for science communication.

Our contributions can be summarized as follows:
\begin{itemize}
\item A conceptual \textit{conversational visualization} framework that couples real-time scientific visualization and natural language processing (NLP) to support groundbreaking new interaction patterns.
\item The tailoring of language models to include domain knowledge. Inclusion of visualization parameters in the NLP prompting and response analysis. Tight coupling of verbal conversations and corresponding visual representations and parameters.
\item A novel \textit{interactive text-to-visualization} method to automatically generate an animation stream providing visual explanation besides the textual information, managing the camera and cutting plane position according to textual input.
\item The \methodname~software prototype as a proof-of-concept application developed with and evaluated by educational experts to automatically explain and demonstrate dense 3D molecular models via conversation.

\end{itemize}
\section{Related work} \label{Related Work}
Our work builds on prior efforts in conversational visualization and visualization for science communication.
\subsection{Conversational Visualization}
The term Conversational Visualization can be interpreted as visualization of conversational data or as interaction with visualization through verbal conversation between the user and the computer. For the remainder of this paper, we adopt the latter definition, which has received a fair amount of attention in recent years.  

Mitra et al.~\cite{FacilitatingConvInteraction} presented a conversational extension to the previous natural language interface NL4DV~\cite{narechania2020nl4dv} so that the user can make a multi-turn dialog with the system and by means of conversation build more complex queries. This work presented the architecture of conversational interaction, which introduced a conversational manager and query resolver submodules, showcased, for instance, in Vega-Lite's visualization authoring and a mind mapping application. Liu et al.~\cite{ADVISor} also proposed a system for visualization synthesis from tabular data. They converted a natural language to vectors presenting the semantic meaning using the BERT model. Vectors served as input to decide the data area, attributes, filter conditions, and aggregation type, which then decided the resulting visual encoding. Luo et al.~\cite{NLtoVISbyNMT} provide a transformer-based sequence-to-sequence neural translation model that translates a natural input related to tabular data into a Vega-Lite grammar syntax that defines the resulting visualization design. Wang et al.~\cite{wang2022towards} contributed an authoring-oriented NLI pipeline and a multi-stage natural language interpreter to parse textual input into a sequence of editing actions. 
Maddigan et al.~\cite{maddigan2023chat2vis}, finally proposed a novel Chat2VIS system to render visualizations from natural language queries. They showed the efficiency of the system by comparing the performance of GPT-3, Codex, and ChatGPT across several case studies.

In contrast to all the mentioned techniques, we leverage large language models, more specifically, GPT-3.5~\cite{openai_chatgpt_2022} and GPT-4~\cite{openai2023gpt4}, for conversational capabilities in our approach. We design a pack-of-bots with a two-level structure. These bots are utilized through prompt engineering and fine-tuning for intent detection and generating visual navigation and exploration commands. This application is focused on challenging 3D, dense, multi-scale models of biological systems. 

\subsection{Visualization for Science Communication}
The use of interactive visualization at museums and science centers is still in its infancy. However, several examples of installations successfully demonstrate scientific findings using visual representations, some of which are interactive. The fact that these installations are placed in a public space places additional requirements on the interaction component, which has to be robust and intuitive. Moreover, visual metaphors must be simplified, and explanations must be embedded in the installation in a non-intrusive way. Recent writings on the state-of-the-art visualization for public spaces provide a comprehensive overview~\cite{Bottinger2020, Ynnerman2020, Rheingans2020}. 

Molecular models can be large, dense, and difficult to interpret, highlighting the need for effective visualization tools. Kou\v{r}il et al. \cite{kouril2021molecumentary} created an adaptable method, Molecumentary, for coupling educational storytelling with engaging visualization. They proposed one self-guided narrative method and one text-to-visualization method. However, Molecumentary did not provide enough interactive ways for the user to explore the model, and the text-to-visualization method could be modified to display the biology information better.

Advances in modern computer graphics make real-time interaction with highly complex 3D models possible. In order to provide the necessary context for non-expert users, a certain level of guidance has to be provided. An early example of studies of interactive visualization at science centers is the plankton table at the Exploratorium in San Francisco~\cite{ma2012plankton,ma2020plankton}.  Ynnerman et al.~\cite{ynnerman:16:inside} used interactive touch tables for the volumetric rendering of CT-scanned mummies at the British Museum. The work included contextualized information tags to tell stories and guide interaction.    

The fusion of exploratory and explanatory visualization was elaborated upon by Ynnerman et al., and examples, including biomolecular visualization and astronomy, were used to show the applicability of the concept~\cite{ynnerman:18:exploranation}.  The exploranation idea has been applied in a wide range of areas, such as the dissemination of results in astronomy~\cite{bock2018openspace,bock2020openspace},  nanoscience~\cite{hoest2020nano}, climate change \cite{besanccon2022exploring}, or molecular dynamics \cite{Brossier:2023:MOL}.
Leveraging the potential of exploranation and its flexibility, Bock et al. \cite{bock2018openspace} have also argued that exploranative systems can be used in museums and science centers with the help of a guide who would take on the storytelling and explanation part of the system while interacting with the system to make it follow their story and explanations. In this context, only a few interaction techniques and systems that would seamlessly assist the guide in their explanations have been developed \cite{Besancon:2021:STAR}.

Current interactive visualization methods for science communication to a wide audience are, except for a few exceptions, still limited to systems (1) of interactive exploration allowing limited manipulations such as selecting color and opacity settings; (2) displaying pre-made animations; (3) relying on external guides and teachers to explain the content while simultaneously interacting with the system. As such, an autonomous system that would interactively provide explanations and exploration capacities would address the needs of a lay audience in science centers while also addressing the possibility of dynamically steering a visualization system as a guide is talking, therefore also closing the gap identified in the literature \cite{Besancon:2021:STAR}. This is what we address with our work.
\section{Design Requirements}
\label{sec:design_requirements}

Our primary objective is to develop an innovative conversational visualization tool that acts as a visual oracle able to assist science communication in public centers. This tool will possess the capability to respond to any inquiry with both spoken and visual answers. It will focus on complex objects, such as the intricate structures within biological models. The target audience for this tool is lay visitors to the science center, who may require assistance in exploring and comprehending the nuances of biological models. Our tool will take on the role of a guide, providing a seamless and intuitive experience. Notably, it will be entirely voice-controlled, eliminating the need for additional input devices. 

To identify the specific needs that this oracle should tackle, we have decided to engage with three (E1, E2, and E3) experts in science communication from Norrköping Visualization Center. These experts have extensive experience in building, evaluating, and using interactive exhibits to explain scientific concepts to lay audiences. Respectively, our expert (E1) has been involved in the making of immersive scientific movies for lay audiences on chemistry concepts, the award-winning ``Chemistry of Life'' movie\footnote{\url{https://visualiseringscenter.se/en/film/chemistry-life}}; the design, use, and study of an interactive visualization system for the Norrköping Visualization Center (E2 and E3). All three experts had more than 20 years of expertise in the study and practice of education, particularly for STEM topics, with a heavy focus on chemistry and biology. All of them have extensive knowledge of and track record in the educational literature and have acted as narrators, guides, or teachers to explain scientific concepts to pupils and lay audiences alike. To elicit the requirements of our oracle for them, we decided to conduct individual interviews with them. In these online interviews that lasted approximately 45 minutes, the experts were asked to imagine what a tool such as our oracle should be able to do and the tasks that it should support. To ensure that the requirements they provide us with would not be hindered by the current limitations of technological solutions, we asked them to imagine that the technology is perfect and can achieve all they can think of. We insisted that the conversation should really be technology-agnostic.

Interestingly, from our discussions with the experts, it appeared that they wanted to distinguish between two different contexts in which our oracle could be used: a \emph{guided context} in which a narrator/teacher/guide with sufficient knowledge of the data provides an audience with explanations while the visualization and oracle are interactively supporting the guide; an \emph{unguided context} in which a lay audience (pupils, visitors) would interact with the visualization on their own and with limited knowledge. In fact, the guided context is mostly a special, easier case than the unguided one. The system only reacts and steers the visualization but does not respond or engage in conversations. Our interviews were video-recorded for error-checking purposes, and all were conducted by a single author. The complete, anonymized, and categorized design requirements (DRs) are in supplementary material section 1. We report below the design requirements that we extracted from the interviews if two experts or more mentioned them. If a design requirement applies only in a specific context, the context is specified.

\DRone: \textit{The oracle should act as a Pilot interactively steering the visualization system}. The knowledge of the data and the oracle's ability to operate view and object manipulations are not enough to provide a good exploranation experience. The oracle should be able to ``fly'' to specific structures of the data on demand based on vocal commands and specific keywords or triggers. It should use its knowledge of the data to ideally show the data or a substructure of it to the audience, including using pre-set visualization encodings or highlighting techniques.

\DRtwo: \textit{The oracle should be able to operate view and visualization-widget manipulation}. Much like one would expect a touch table in a science center to allow for simple view and object manipulations of a dataset, the oracle should conduct such operations. This allows users to explore and focus on specific parts of the data, therefore maintaining an interactive experience. The manipulation of cutting planes or other visualization widgets may help analyze the internal structure of data on their own.

\DRthree: \textit{The oracle should have contextual awareness for the presented data}. This is a fundamental requirement that other design requirements build onto. The oracle should have in-depth knowledge about the data displayed by the system such that it can ``explain something on its own as well'', ``complement the information that the visualization is giving'', or ``answer basic or complicated queries/questions.''

\DRfour: \textit{The oracle should be able to change representation modes and metaphors}. Molecular datasets can be particularly complex, and the ideal representation or visual encoding may depend on the part of the data that is being focused on. As such, the oracle must be able to change different visual parameters (e.g., transparency, emphasis) or interactively select representation modes (e.g., Van der Waals spheres/ribbons) to facilitate the audience's learning experience.

\DRfive:  \textit{The oracle should be intuitive and offer initial guidance for exploration}. The oracle should not require a long tutorial (which is known to discourage exploration and learning in science centers) to be used. It should maybe offer some simple hooks or prompts to initiate an interactive learning experience. ``Maybe a command that guides with a `help' or `tell me how to use your functions' '' (E1) or directly ``some expected outcomes when interacting with the system'' (E3). This is particularly true in the \emph{unguided context} since guides are not supposed to need an introduction to the functionalities.

\DRsix: \textit{The oracle should adapt to its audience and their knowledge.} The oracle should be able to deduce from its interaction with the audience their level of understanding of the data/concept and adapt its explanations, representations, and words to match the identified expertise. It should also adapt its interactivity levels to provide more storytelling and information to a naive audience than to an audience that seems to have more knowledge. This is particularly true in the \emph{unguided context}.

\DRseven: \textit{The oracle should also analyze and adapt to group dynamics.} The oracle should be able to analyze the interactions between the audience and guide and change the content and modalities used depending on the signals coming from the interactions between visitors and guides. It should also make sure not to produce redundant content (or modalities) with the guide to avoid overloading the audience. This is particularly true in the \emph{guided context}.

While all of these requirements have been highlighted by our experts to foster a better educational and conversational outcome in public centers, the last two requirements would require technological and analytical capabilities that would go beyond the scope of this paper. We, therefore, only focus on the first five design requirements for the remainder of this paper and discuss in \autoref{sec:discussion} how the last two could impact future work and learning opportunities in science centers.
\section {Method Overview } \label{method}

Our aim is to develop an oracle combining the capabilities of advanced conversational visualization, addressing the design requirements 1-5.

The conversational abilities of large language models, specifically GPT-3.5 and GPT-4, have reached a maturity where they convey information on general topics with relatively high reliability. When prompted with detailed descriptions of tasks, these transformer networks produce highly accurate responses. Therefore, we consider GPT-4 as suitable for the role of an oracle that we can integrate with visualizations. The model can instantaneously provide an answer when the user asks a question related to the object of interest. However, what should be displayed in the visualization to accompany the answer may not be immediately clear. To address this, an algorithm can analyze the user input and attempt to determine the appropriate visualization to display. An even better solution is to request that GPT-4 provide a response indicating what the user wants to see, allowing a more accurate visualization experience tailored to users' needs.
\begin{figure*}[ht]
  \centering
  \includegraphics[width=0.9\linewidth]{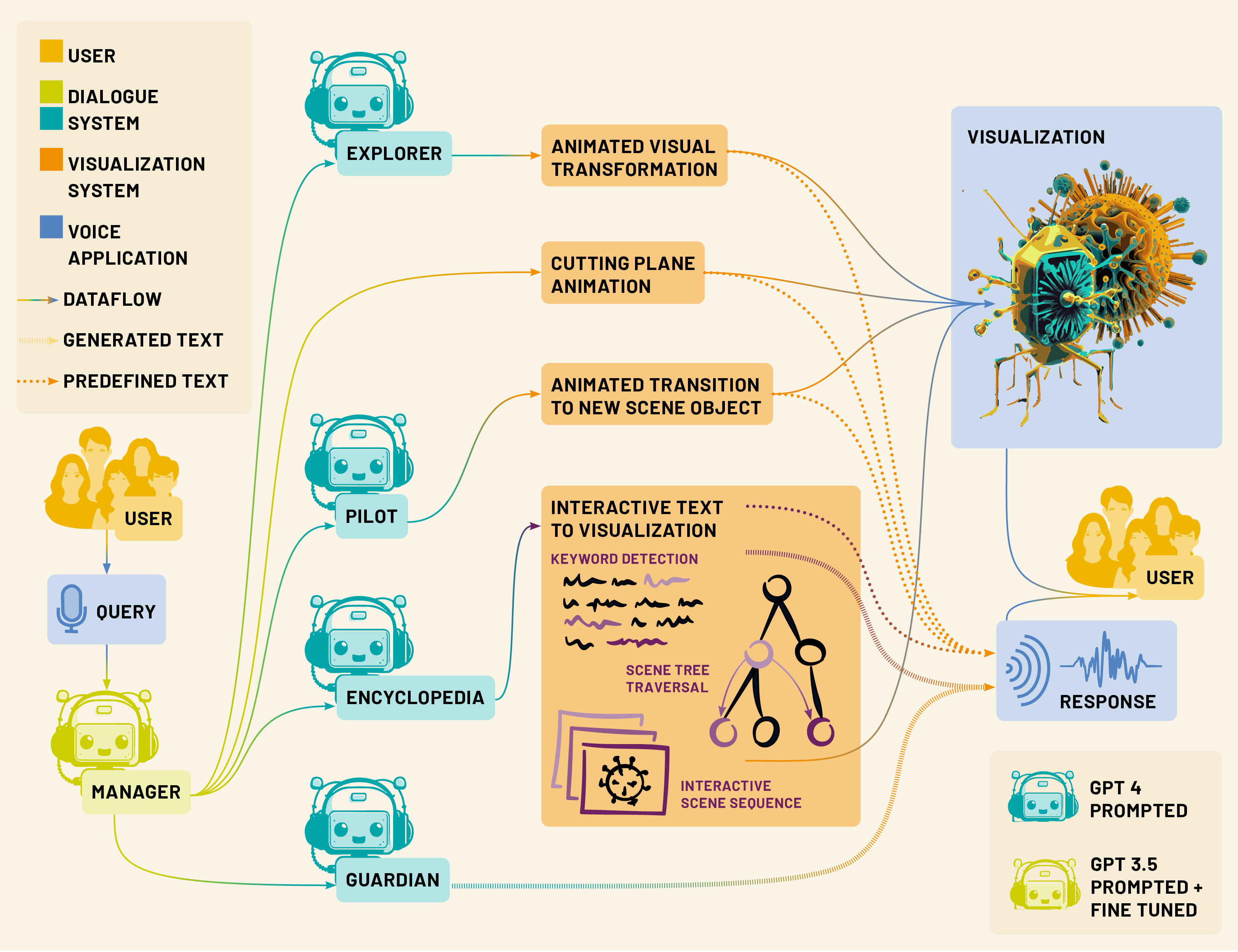}
  \caption{%
\textbf{Overview of the \methodname~framework.} The dialogue system begins with a user's speech query. It uses a ``pack-of-bots'' architecture to process this query. The system either answers questions or follows instructions, which are then given to a visualization system.}
   
  \label{fig:sketch_architecture_detail}
\end{figure*}
Our method is based on the use of bots. To achieve the above design requirements, we have identified two distinct tasks: providing knowledge and informing the visualization of what to display. Both of these tasks require bots with well-designed prompts to obtain accurate answers. However, a single bot cannot be effectively prompted for multiple distinct roles. To address this, we introduce the pack-of-bots strategy in our dialogue system (depicted on the left side of \autoref{fig:sketch_architecture_detail}), which utilizes a set of bots, each prompted or fine-tuned for a specific role in which it excels. Our pack-of-bots employs a two-level design. The first level is formed by a \textit{Manager} bot, which classifies the intent of the user's request from the visual oracle. The user might ask a question about the content of what they currently see, which is passed on to the \textit{Encyclopedia} bot. If the user wants to request minor visualization modifications such as changing a viewpoint or manipulating a cutting plane, the \textit{Manager} bot passes on the task to the \textit{Explorer} bot or the \textit{Cutting Plane} process. If the user requests to see another object or see the scene from a different level of detail, the \textit{Manager} bot passes the task to the \textit{Pilot} bot. If the user statement is irrelevant to the current scene, the \textit{Guardian} bot guides the user away from irrelevant topics. The text output of the \textit{Encyclopedia}  and \textit{Guardian} bots is synthesized into speech directly. The \textit{Exploration} and the \textit{Pilot} bot both extract the operation-specific intents from the forwarded user query and translate them into our internal instruction syntax, which is then processed by the visualization system. The \textit{Cutting Plane} process is triggered directly by the \textit{Manager} bot and skips the additional internal intent-detection step. The \textit{Narration} generator is designed to transfer our internal instructions to human-sounding answers. We manually design several responses for each type of task and select randomly every time, which will be introduced later in \autoref{Data}. We summarize our query processing approach in algorithm \autoref{alg:queryprocessing}. The interactive text-to-visualization module receives the output from the \textit{Encyclopedia} bot, creating an interactive flythrough that navigates through the objects mentioned in the output.

These objects are described in a semantic hierarchy tree consisting of multiple levels of details. Our visualization system (depicted in the middle of \autoref{fig:sketch_architecture_detail}) traverses this hierarchy of scene elements. For example, a model of a T4 bacteriophage could be decomposed into the head and tail (shown on the labels of \autoref{fig:teaser}). The head could be further decomposed into the protal protein, major capsid protein, etc. (shown on the labels of \autoref{fig:scene} (a). The interactive text-to-visualization method contains three components: \textit{keyword detection}, \textit{scene tree traversal}, and \textit{scene sequence}. The \textit{keyword detection} aims to detect all the elements in the text. The \textit{scene tree traversal} method returns an ordered \textit{scene sequence} based on the hierarchical scene tree structure. We define several types of \textit{scene}, and each scene generates several frames to compose an animation.

\begin{algorithm}[t]
\caption{Query processing} \label{alg:queryprocessing}
\begin{algorithmic}[1]
\State $type \gets$ managerBot($query$)
\Switch{$type$}
     \Case{$Encyclopedia$:}
         \State $reply \gets$EncyclopediaBotGPT-4($query$)
         \State $narration \gets$narrationGenerator($type$,$query$)
         \State interactive text-to-visualization($reply$) 
         \EndCase
     \Case{$Explorer$:}
        \State $\vec{t} \leftarrow $ExplorerBotGPT-4($query$)
        \State $narration \gets$narrationGenerator($type$,$\vec{t}$)
        \State visualTransformation($\vec{t}$)
        \EndCase
     \Case{$Pilot$:}
         \State $reply \gets $PilotBotGPT-4($query$)
         \State $narration \gets$narrationGenerator($type$,$reply$)
         \State transationAnimation($reply$)
         \EndCase
     \Case{$Cutting\ Plane$:}
        \State $narration \gets$narrationGenerator($type$,$reply$)
         \State cuttingPlainAnimation($query$)
         \EndCase
     \Case{$Guardian$:}
          \State $reply \gets $GuardianBotGPT-4($query$)
         \State GuardianProcess($reply$)
         \EndCase
\EndSwitch

\end{algorithmic}
\end{algorithm}
Technical details are given in \autoref{dialoguesystem} and \autoref{vissystem}, while the implementation details are outlined in \autoref{result}.

\section{Dialogue system}\label{dialoguesystem}
The dialogue system parses the user's input, provides knowledge, or extracts concrete machine-readable instructions. In this section, we discuss how we implement the \textit{pack-of-bots} architecture based on the design requirements in detail. 

\subsection{Prompt-based fine-tuned Classifier} \label{classifier}
\begin{figure}[ht] 
\centering
    \includegraphics[width=0.7\linewidth]{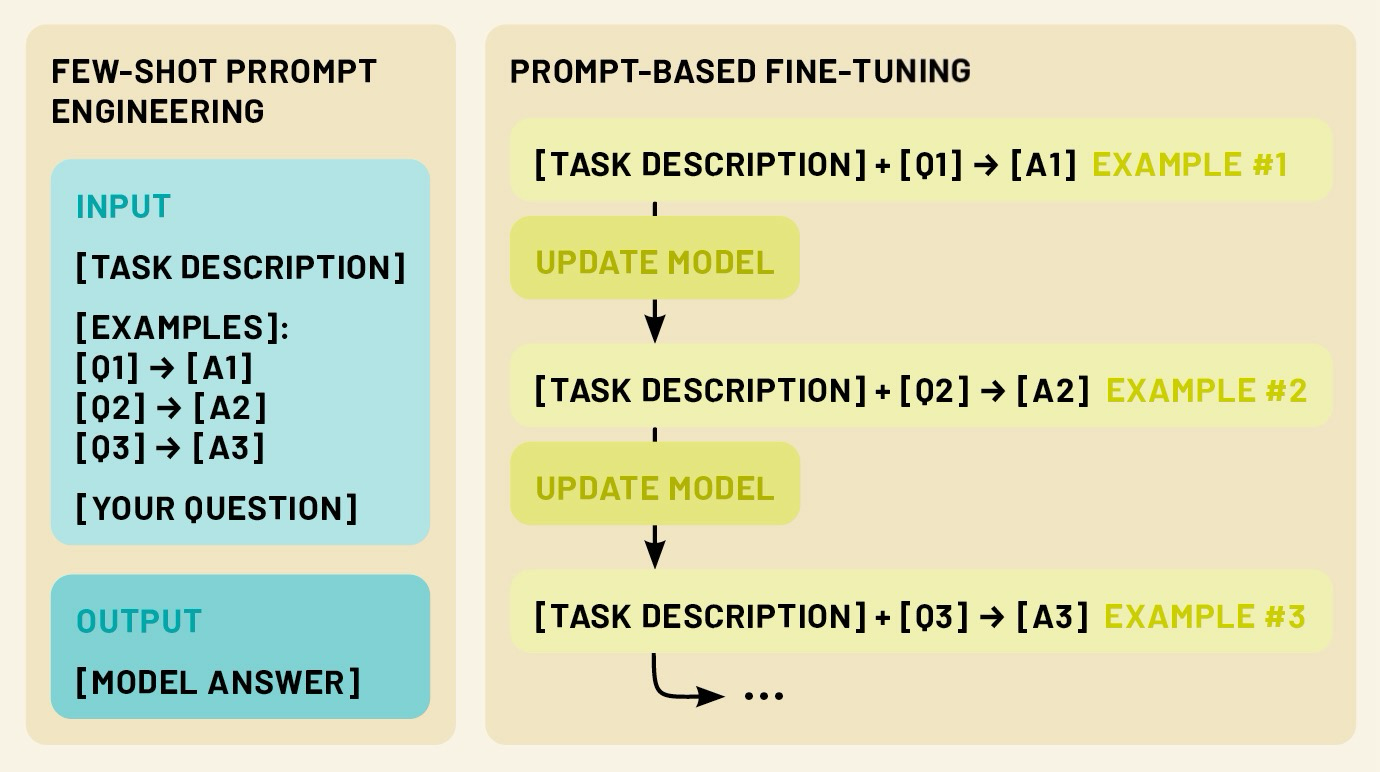}
    \caption{%
    \textbf{Few-show prompt engineering and prompt-based fine-tuning.} Few-shot prompt engineering enables direct output acquisition without altering the model. Conversely, prompt-based fine-tuning updates the model through multiple steps.
    }
    \label{finetuneandprompt}
\end{figure}

As described in \autoref{method}, we utilize LLM to determine the appropriate visualization to display. The first step of our visual oracle method is that the \textit{Manager} bot runs an intent analysis task to ensure that different types of commands are assigned to the appropriate bots, turning it into a multi-classification problem.  While large language models are capable of understanding general semantics, they still require carefully crafted prompts to complete specific tasks, i.e., prompt engineering. Prompt engineering aims to modify the input fed into the model to achieve better performance without changing the model's weights. In our method, we follow the few-shot learning setting for all the bots except \textit{Manager} bot, where the model is provided with the task description and a few examples of the task at inference time, but there are no weight updates of the model. For instance, to enable a large language model to distinguish the action of rotation or zoom, a detailed description of these actions must first be composed as a task description. Subsequently, the model must be told a few sample examples, such as ``Turn right$\rightarrow$rotation'' and ``Move closer$\rightarrow$zoom'' along with the actual query you want, such as ``Show the top of this object.''. This approach is illustrated on the left-hand side of \autoref{finetuneandprompt}. 

Our experience has shown that prompt engineering alone is insufficient to solve our intent analysis task. One reason is that we cannot feed enough examples to the model due to token limitations, i.e., the model cannot process too much text in a single request at inference time. Longer input results in longer processing time. Another reason is that our task is too complex in the semantic aspect, and prompt engineering alone may not be sufficient to capture the full range of variation in the data. Therefore, we employ a prompt-based fine-tuning method, which is demonstrated on the right-hand side of \autoref{finetuneandprompt}. This approach involves fine-tuning the large language model with the dataset, where the data is fed together with the task description as the prompt, and the model weights are updated at each step. Fine-tuning has demonstrated better performance than prompt engineering but takes more time and resources to collect the data set and fine-tune the model.

More specifically, to address the design requirements, we define five classes (\textit{Pilot}, \textit{Cutting Plane}, \textit{Explorer}, \textit{Encyclopedia}, and \textit{Guardian}) and fine-tuned GPT-3.5 model using our own dataset. We discuss how we identify each class in \autoref{chatbots} and how we build the dataset in \autoref{largelanguagemodelselection}.

\subsection{Prompted Chatbots} \label{chatbots}

The prompt engineering is consistently applied to the remaining bots following the few-shot learning paradigm. Specifically, we prompt each bot with a set of rules for the given task and a few examples. Remarkably, the bots can accomplish each task accurately without requiring further fine-tuning. In practice, we employ GPT-4 as the model to be prompted. We list a few examples of which bot the query will be assigned to and the corresponding output in \autoref{tab:examples}.

To tackle design requirement 1 (steering), we need a bot that can intuitively discern the user's preferences regarding the aspects of the molecular model. Additionally, given the diverse scales present within the molecular model, the ability to seamlessly modify the scale is paramount. To enhance the user experience, we introduce a reset functionality, allowing users to restore default camera positions, cutting plane settings, and more. Furthermore, we empower users with a unique ``look back'' function, enabling them to retrace their steps and revisit previously explored nodes, thereby facilitating a more comprehensive exploration of the molecular model. Therefore, we design the \textit{Pilot} bot to distinguish between the four defined classes, which refer to \textit{node navigation}, \textit{scale change}, \textit{reset}, and \textit{return back}.  We find that we get better results by introducing the rules as a decision tree method with a list of conditional statements. This stems from our classification based on our examples, which could fit into multiple categories. For example, \textit{go back to the Capsid} will fit both the \textit{node navigation} class and the \textit{return back} class. In case of the direct mentioning of an object, we want it to opt for \textit{node navigation} class, regardless of whether it is the last or second-to-last element we looked at. The examples include straightforward examples as well as equivocal ones. The reply is a single digit indicating the command classification from 1 to 4. The full prompt can be found in supplementary material section 2.

The \textit{Pilot} bot is primarily engineered to provide users with a visual representation of the external biological structure. Nevertheless, it is equally crucial to make detailed information about the interior components of the complex biological structure available to users. Consequently, we have introduced a \textit{Cutting Plane} class within the \textit{Manager} bot's design. This design allows for dynamic manipulation of the cutting plane, enabling users to explore the intricate interior of the complex structure. It's worth noting that since there are no anticipated future processing requirements, we have not designed any additional bots to process user commands; instead, these cutting plane commands are directly transmitted to the visualization system for immediate execution.

\begin{table}[ht]
\centering
\begin{tabular}{ccllc}
\toprule
Class              & \multicolumn{3}{c}{Query}                                                                                   & Output                                                        \\ 
\midrule
\multirow{2}{*}{Explorer} & \multicolumn{3}{c}{\begin{tabular}[c]{@{}c@{}}I want to see the right \\ side of this object.\end{tabular}}  & \{1,90,0,0\}                                                  \\ \cmidrule{2-5} 
                          & \multicolumn{3}{c}{\begin{tabular}[c]{@{}c@{}}It's too far. I want \\ it up close.\end{tabular}}          & \{2,0,0,0\}                                                 \\ \midrule
\multirow{4}{*}{Pilot}    & \multicolumn{3}{c}{Show me the capsid.}                                                                     & \begin{tabular}[c]{@{}c@{}}node \\ navigation\end{tabular}    \\ \cmidrule{2-5} 
                          & \multicolumn{3}{c}{Go back to the start.}                                                                   & reset                                                         \\ \cmidrule{2-5} 
                          & \multicolumn{3}{c}{Go up a level.}                                                                          & \begin{tabular}[c]{@{}c@{}}scale \\ modification\end{tabular} \\ \cmidrule{2-5} 
                          & \multicolumn{3}{c}{Show me the last thing again.}                                                           & return back                                                   \\ \midrule
Encyclopedia              & \multicolumn{3}{c}{\begin{tabular}[c]{@{}c@{}}What is the \\  matrix protein?\end{tabular}} & \begin{tabular}[c]{@{}c@{}}generated \\ text\end{tabular}     \\ \midrule
Guardian                  & \multicolumn{3}{c}{\begin{tabular}[c]{@{}c@{}}Please play \\ music for me.\end{tabular}}                   & \begin{tabular}[c]{@{}c@{}}generated \\ text\end{tabular}     \\ \bottomrule
Cutting Plane                & \multicolumn{3}{c}{\begin{tabular}[c]{@{}c@{}}Please show me the\\ interior objects.\end{tabular}}                   & \begin{tabular}[c]{@{}c@{}}cutting plane \\ modification\end{tabular}     \\ \bottomrule
\end{tabular}
\caption{\textbf{Examples of the query and output.} For each query, the output format is controlled by the corresponding class.}
\label{tab:examples}
\end{table}

To address design requirement 2 (view), a dedicated bot is necessary to extract visual parameters from user commands. In response to this requirement, we have designed the \textit{Explorer} bot. This specialized bot is responsible for parsing and processing user requests, converting them into specific visual commands for execution. It is directly prompted by the underlying task of extracting zoom and rotation. The set of rules we designed summarizes the task, defining the return format $ \{\left[zoom\ factor\right],\left[yaw\right],\left[pitch\right],\left[roll\right]\}$, giving the option to return the default transformation values if no visual instructions are detected, and the order to interpret inexact values as seen fit (e.g., ``a little''). They are also derived from GPT-4 inquiries for more context (always reminding it of the initial view direction being $\vec{v} = (0, 0, -1)$), as well as designed to avoid mistakes of previous iterations of prompts (reminding it, for example, of adjacent faces and that we are looking at it from the front).

The \textit{Encyclopedia} bot operates with more flexibility compared to the two preceding bots. Its primary purpose is to provide knowledge and address design requirement 3 (context). To enhance its collaboration with the interactive text-to-visualization method, we have integrated model-specific keywords directly into the prompts and provided clear instructions on their usage.

To ensure that our \textit{Encyclopedia} bot focuses solely on the domain at hand, in our primary example, the molecular-related questions, we take the additional step of assigning the \textit{Guardian} bot the task of processing any ordinary questions that fall outside the scope of our visual oracle. The \textit{Guardian} bot is prompted to provide truthful answers to such questions and redirect the user's attention to our 3D molecular model at the appropriate time. This arrangement is designed to prevent any unwanted interaction between the \textit{Encyclopedia} bot and the 3D model, thus preserving the integrity of our visual oracle.

\section{Visualization System}\label{vissystem}
In this section, we will discuss the step-by-step process involved in the scene tree setting and defining scenes for our visualization system. Specifically, we will focus on the \textit{interactive text-to-visualization} method, which is responsible for converting textual input into the corresponding animation. Throughout this discussion, we will highlight the key considerations and challenges involved at each stage of the process.

\subsection{Scene Tree Definition} \label{hierarchy tree}

\begin{figure}[ht] 
    \centering
    \includegraphics[width=0.7\linewidth]{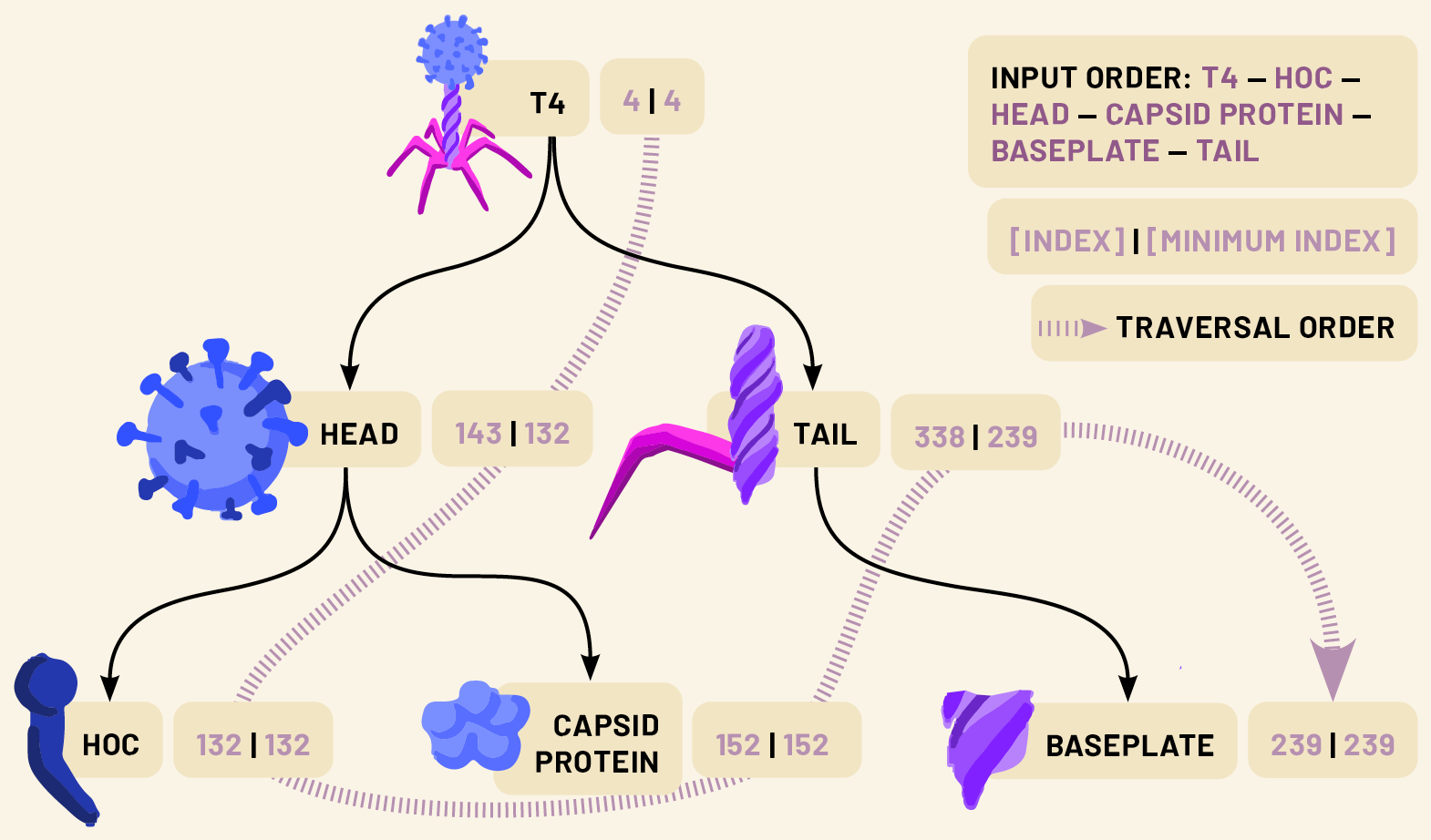}
    \caption{%
    \textbf{The interactive text-to-visualization method is demonstrated in the scene tree.} 
    Minimum index values are updated based on the index value of each node. Then, a traversal list is generated based on the minimum index values. }
    \label{fig:Scene Tree}
\end{figure}

We base our description on the molecular domain example. Three-dimensional molecular models have multi-scale and multi-instance properties within the dense environment, making the design for exploring the inside challenging. To better express the hierarchy structure of the 3D molecular model, we adopt a tree data structure, illustrated in \autoref{fig:Scene Tree}. In our cases, each node represents one ingredient type, containing its name, label, parent relationship, instances information, and $minimumIndex$ value. We also add a one-sentence description for each node, describing its components and functions.
\begin{figure*}[ht]
    \centering
    \subfigure[Overview scene of the \textbf{Head} node]{
    \includegraphics[width=0.25\linewidth,height=4.5cm]{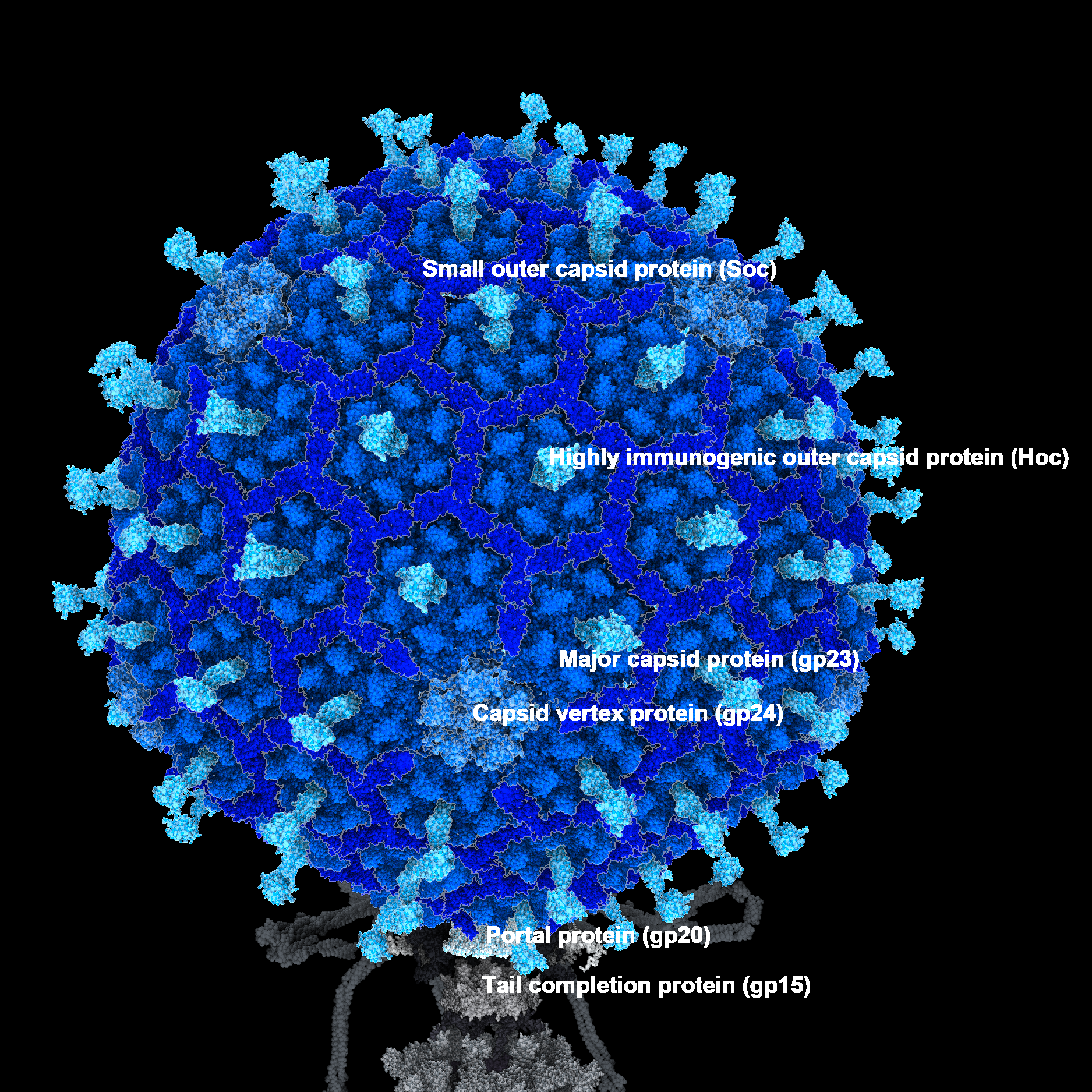}}
    \subfigure[Focus scene of the \textbf{portal protein}]{
    \includegraphics[width=0.25\linewidth,height=4.5cm]{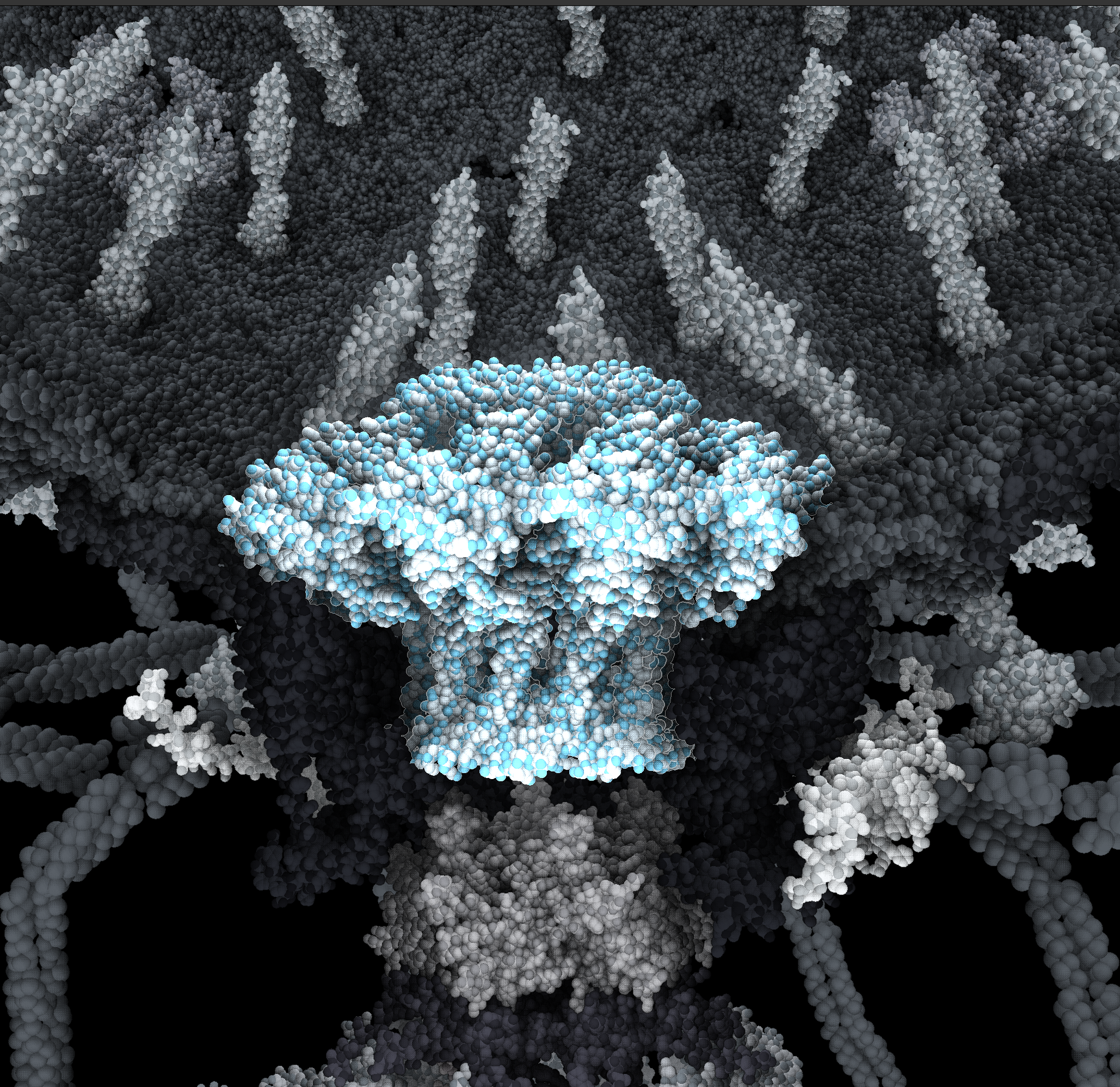}}
    \subfigure[Cutting plane scene of the \textbf{Head} node]{
    \includegraphics[width=0.25\linewidth,height=4.5cm]{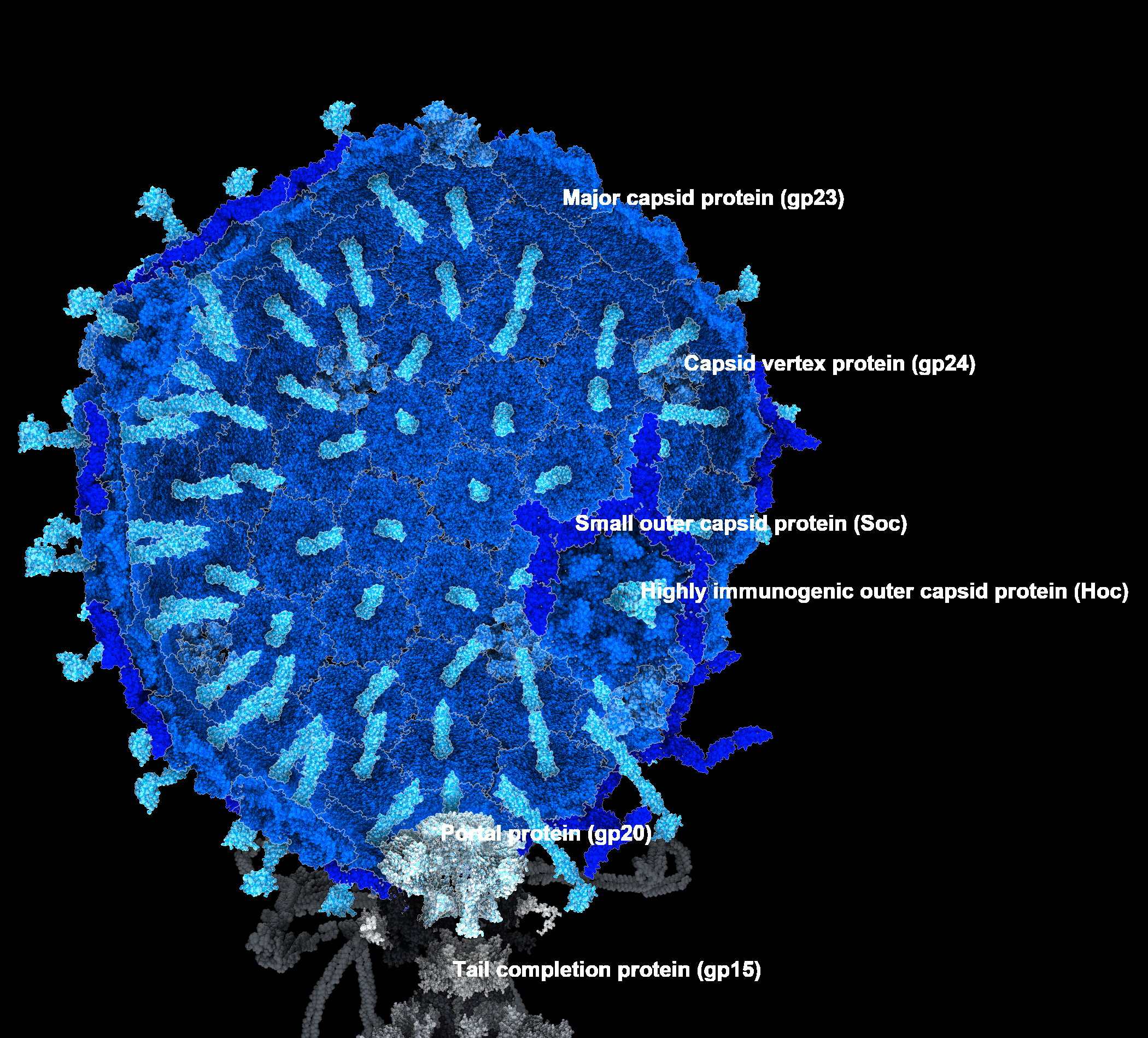}}
    \caption{\textbf{Overview scene (a), focus scene (b), and cutting plane scene (c).} The overview scene shows the external component labels and spatial information, while the focus scene illustrates the structural details. The cutting plane scene displays the internal components.}
    \label{fig:scene}
\end{figure*}

\subsection{Timeline and Scenes} \label{scene and timeline}
We utilize the \textit{queue} structure to implement the \textit{timeline} following the first-in-first-out approach. The timeline stores the scene sequence and displays the new scene from the front of the queue after the current scene finishes. The current scene finish status is controlled by the speech completion signal and animation completion signal.

To address design requirement 4 (style), we need the system to modify the visualization representation based on the focusing ingredient. Following the Molecumentary\cite{kouril2021molecumentary} framework, we maintain the two established types of scenes: \textit{Focus}, \textit{Overview}. In addition to these, we introduce two new scene types: \textit{Cutting Plane} and \textit{Speech Only}.

The \textit{Focus} scene, depicted in \autoref{fig:scene} (b), is dedicated to presenting in-depth information about the specific structure type. It achieves this by identifying the closest instance of the type near the current camera position and moving the camera in closer proximity. Furthermore, a subtle rotation animation is generated to offer a different perspective. In contrast to Molecumentary's approach, which directly reveals the internal components of the \textit{Overview} scene object, we think the external spatial information and structure display are also important; therefore, we redesign the \textit{Overview} scene, as illustrated in \autoref{fig:scene} (a). This scene showcases the view of the nearest object to emphasize its external components. To enhance clarity, all internal components are highlighted to facilitate discrimination. Here, we employ the bounding sphere technique \cite{kouvril2020hyperlabels} to reposition the cutting plane to the sphere's border. We introduce the \textit{Cutting Plane} scene, shown in \autoref{fig:scene} (c), to display the interior components of the object. Similar to the \textit{Overview} scene, we also highlight the internal components to clarify the object's scope.

We have implemented transitional sentences within a dedicated \textit{Speech Only} scene to enhance user guidance and maintain a coherent and fluid conversation. We provide a subtle rotation animation of the current object to prevent the current scene from becoming monotonous.

\subsection{Interactive Text-to-Visualization} 
\label{texttovis}

Upon receiving the textual output from the \textit{Encyclopedia} bot, we require an interactive text-to-visualization method to parse the text and generate appropriate animations. However, the current text-to-visualization method in Molecumentary exhibits four significant drawbacks, hindering meeting our requirements. First, single keyword limitation: the current method only captures the first keyword within a sentence, even if multiple relevant keywords are present in the text. This limitation results in a loss of crucial textual information, particularly when the text contains multiple keywords that are integral to the visualization. Second, lack of sentence-keyword alignment: the existing method treats the entire sentence as the description for the selected keyword without considering whether the sentence contextually aligns with the keyword. Third, ignore scene tree information: the method overlooks the essential scene tree information, which plays a pivotal role in conveying the structural hierarchy of the content to users. This omission prevents users from obtaining a comprehensive understanding of the visual content. Fourth is information overload: The method simultaneously displays the animations for many keywords. This approach can overwhelm users with excessive information, diminishing their ability to freely choose their object of interest within the visualization.

We propose the \textit{interactive text-to-visualization} method to overcome these drawbacks. The method is also a keyword-based algorithm, but it supports the exploration in a more interactive way and considers the scene tree information, exploring objects from the root node to leaf nodes. We have structured the scene tree into distinct paths, each defined by the shortest route from the root node to a specific leaf node. To determine the appropriate animation sequence, we compare the order of appearance in the text for each path. The detailed algorithm for this process is provided in algorithm \autoref{alg:Node Sorting} and algorithm \autoref{alg:text to visualizaiton}.

\begin{algorithm}[h]
\caption{Node Sorting} \label{alg:Node Sorting}
\begin{algorithmic}[2]
\Procedure{updateMinimumIndex}{node,nodeMap}
\State \textbf{if}\ !$node$\ \textbf{return}
\State \textbf{if}\ nodeMap.contains($node$)
\State \ \ \ \  \textbf{if} nodeMap[node]$<\ node.minimumIndex$
\State \ \ \ \ \ \ \ \ $node.minimumIndex$=nodeMap[node]
\State \textbf{for each}\ $chilNode$\ \textbf{in}\ $node.children$
\State \ \ \ \ updateMinimumIndex($chilNode$, $nodeMap$)
\State \ \ \ \ $node.minimumIndex$=min($chilNode.minimumIndex$,
\State \ \ \ \ \ \ \ \ \ \ \ \ \ \ \ \ \ \ \ \ \ \ \ \ 
 \ \ \ \ \ \ \ \ \ \ \ \ $node.minimumIndex$)
\EndProcedure
\Procedure{sortNodes}{$node$, $nodeMap$,$result$}
    \State \textbf{if}\ !$node$\ \textbf{return} 
    \State \textbf{if}\ $nodeMap$.contains($node$)
    \State \ \ \ \ $result$.add($node$)
    \State \ \ \ \ nodeMap.remove($node$)
    \State $chilNodeList \gets$ sort children nodes by minimumIndex
    \State \textbf{for each}\ $chilNode$\ \textbf{in}\ $chilNodeList$
    \State \ \ \ \ sortNodes($chilNode$, $nodeMap$, $result$)
\EndProcedure
\State $input \gets$ input text
\State Map$<$node,index$>$ $nodeMap $
\State $nodeMap \gets$ keywordDetection($input$)
\State updateMinimumIndex($rootNode$,$nodeMap$)
\State sortNodes(root,nodeMap,result)
\State \textbf{return} result
\end{algorithmic}
\end{algorithm}

We first design the \textit{Node Sorting} algorithm to parse the text and get an ordered node list, shown in algorithm \autoref{alg:Node Sorting}. Firstly, we parse the input text to detect all occurrences of node names taken from our scene tree. We then store both the node and its first occurrence position as $index$ in the text into a map structure called $nodeMap$. The nodes are structural types that build up the scene, as shown in \autoref{fig:Scene Tree}. Next, we update the $minimumIndex$ value of each node in the scene tree based on the $index$ value and the $updateMinimumIndex$ function. The $minimumIndex$ is the metric to compare different paths of the whole scene tree. We perform this update from the bottom of the tree to the top, with the $minimumIndex$ of each node being determined by the minimum $minimumIndex$ value of itself and all its child nodes. Finally, we apply a depth-first search to traverse the scene tree and generate a sorted list of nodes, utilizing the $minimumIndex$ value as the comparison metric for all child nodes. This step is summarized in the $sortNodes$ function.

To illustrate the concept, consider the input text: ``The T4 bacteriophage is a complex virus that infects bacterial cells. It is composed of multiple protein structures, examples being HOC in the head, or capsid proteins, which protect the genetic material of the virus. Structures like the baseplate, which attaches to the host cell's surface and injects the viral DNA, are located in the tail of the Virus''. We illustrate the entire process in Figure~\ref{fig:Scene Tree}. We first compute the $index$ of each node. For example, the $index$ of `head' is 143 because 142 characters are before the ``head'' word in the text. By utilizing the $updateMinimumIndex$ method, we compute the $minimumIndex$ of every node and get a sorted list of \textit{T4--head--HOC--capsid protein--Tail--baseplate} based on the $minimumIndex$ values.

\begin{algorithm}[ht]
\caption{Interactive Text to Visualization} \label{alg:text to visualizaiton}
\begin{algorithmic}[3]
\Procedure{addToTimeline}{$node$}
\State \textbf{if} hasChildren($node$)
\State \ \ overviewScene(node)
\State \textbf{else}
\State \ \ focusScene(node)
\EndProcedure
\State node $\gets$ parse question text
\State addToTimeline(node)
\State nodeList $\gets$ nodeSorting(answer) 
\State \textbf{While}(continue\_explore)
\State \ \ \ \ select \textbf{n} options from nodeList
\State \ \ \ \ node $\gets$ wait user selects one node
\State \ \ \ \ addToTimeline(node)
\end{algorithmic}
\end{algorithm}
Our \textit{interactive text-to-visualization} algorithm, detailed in algorithm \autoref{alg:text to visualizaiton}, takes two main inputs: the user's $question$ and the $answer$ generated by LLM. The whole algorithm starts with parsing the $question$ text. We select the deepest node from the scene tree and push the node into $timeline$ by the $addToTimeline$ function. The $answer$ is set as the speech. Then the \textit{Node Sorting} algorithm parses the $answer$ and returns an ordered node list, denoted as the $nodeList$. We select the first $n$ node from the $nodeList$ and ask users which one they want to explore. In practice, the $n$ is set to 2. To solve the lack of sentence-keyword alignment problem, we prepare the description of each node as the speech. A collaborator (who is also a co-author of this article) with expertise in both bioscience and computer science, lead the modeling of bacteriophage T4 and SARS-CoV-2. This collaborator establishes the node hierarchies and descriptions based on a comprehensive biology literature review~\cite{yap2014structure, yao2020molecular}. Each description is constrained to 25 words.

The interactive exploration process persists until the algorithm detects that the user intends to conclude the interaction. This determination is made based on whether the user explicitly signals the end of the exploration or inquires about other objects.

Our interactive text-to-visualization algorithm has three main advantages: firstly, it effectively identifies all nodes mentioned in the text, ensuring no loss of information. Importantly, it doesn't require users to view all nodes, allowing for flexibility. Secondly, the algorithm conveys hierarchy information by presenting nodes in a specific order, guiding the exploration from distant to near and from larger scale to smaller scale. This enhances the user's understanding of the structure. Thirdly, the algorithm ensures that the speech and the currently displayed scene object are closely aligned.

\newcommand{\optim}{0.25}
\section{Implementation}\label{result} 
\begin{figure}[ht]
  \centering
  \subfigure[T4 – initial view]{
  \centering
    \includegraphics[width=\optim\linewidth, height=\optim\linewidth]{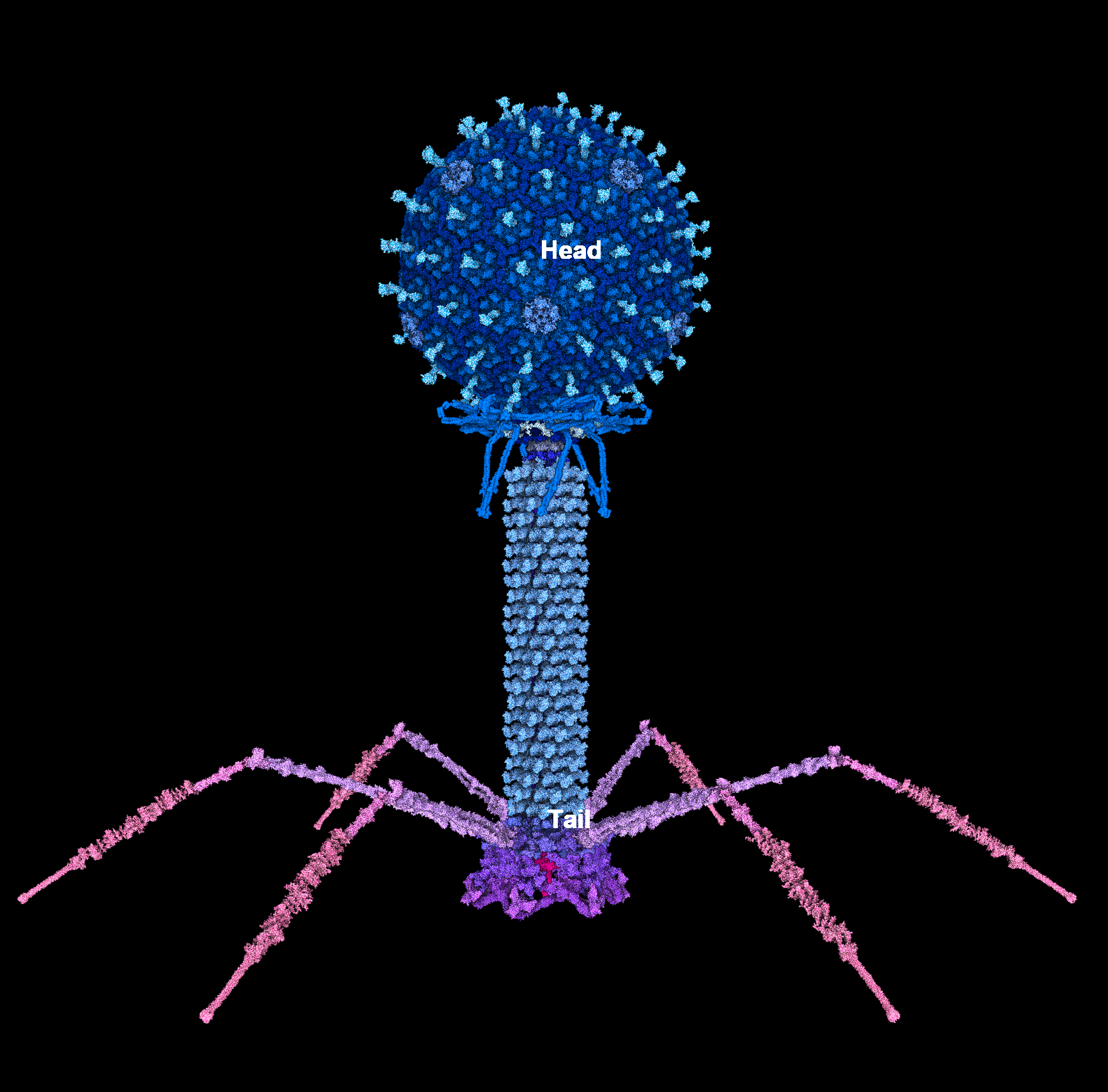}
}
  \subfigure[Show a bottom view.]{
  \centering
    \includegraphics[width=\optim\linewidth, height=\optim\linewidth]{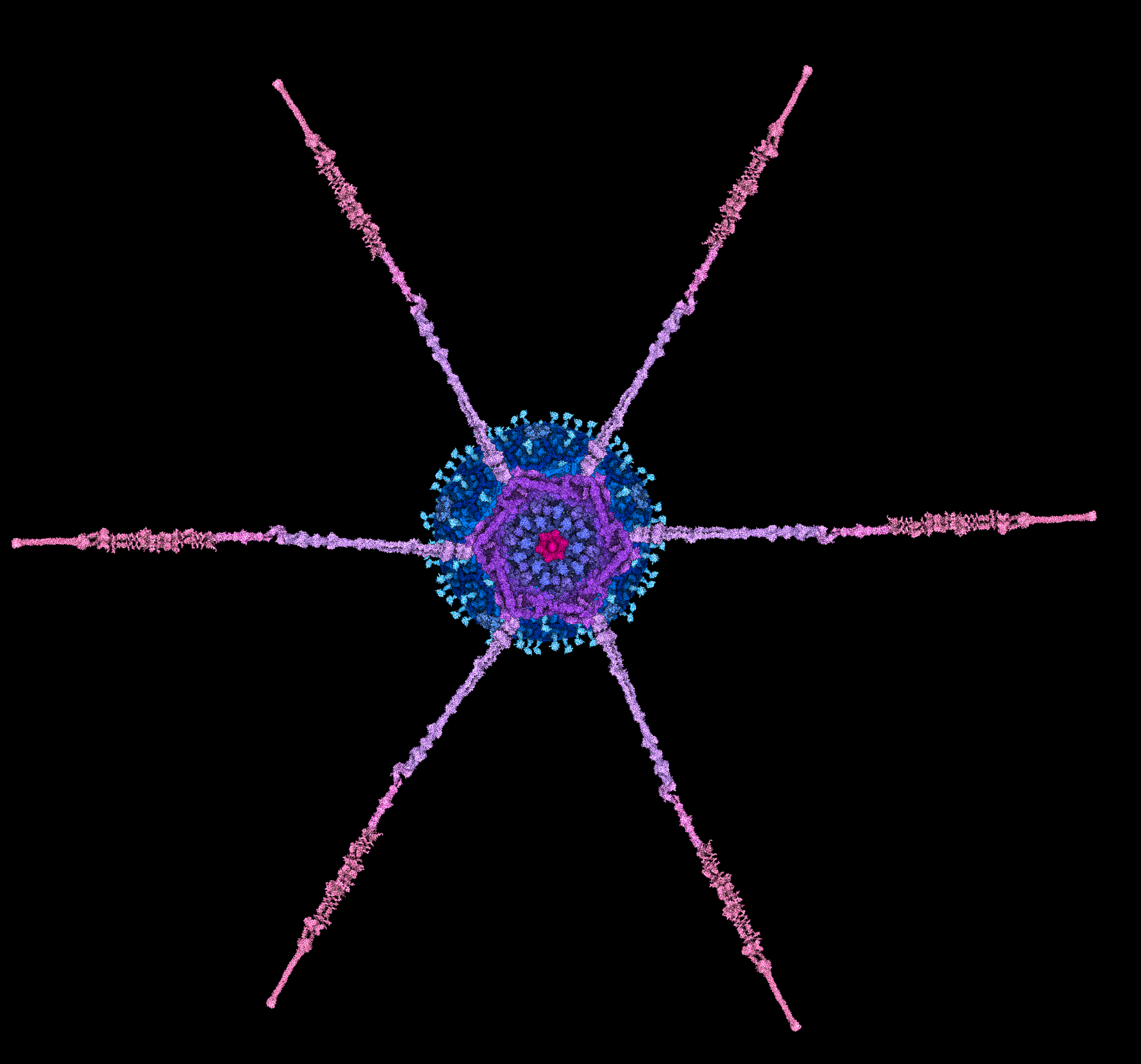}
  }
  \subfigure[Show the baseplate.]{
  \centering
    \includegraphics[width=\optim\linewidth, height=\optim\linewidth]{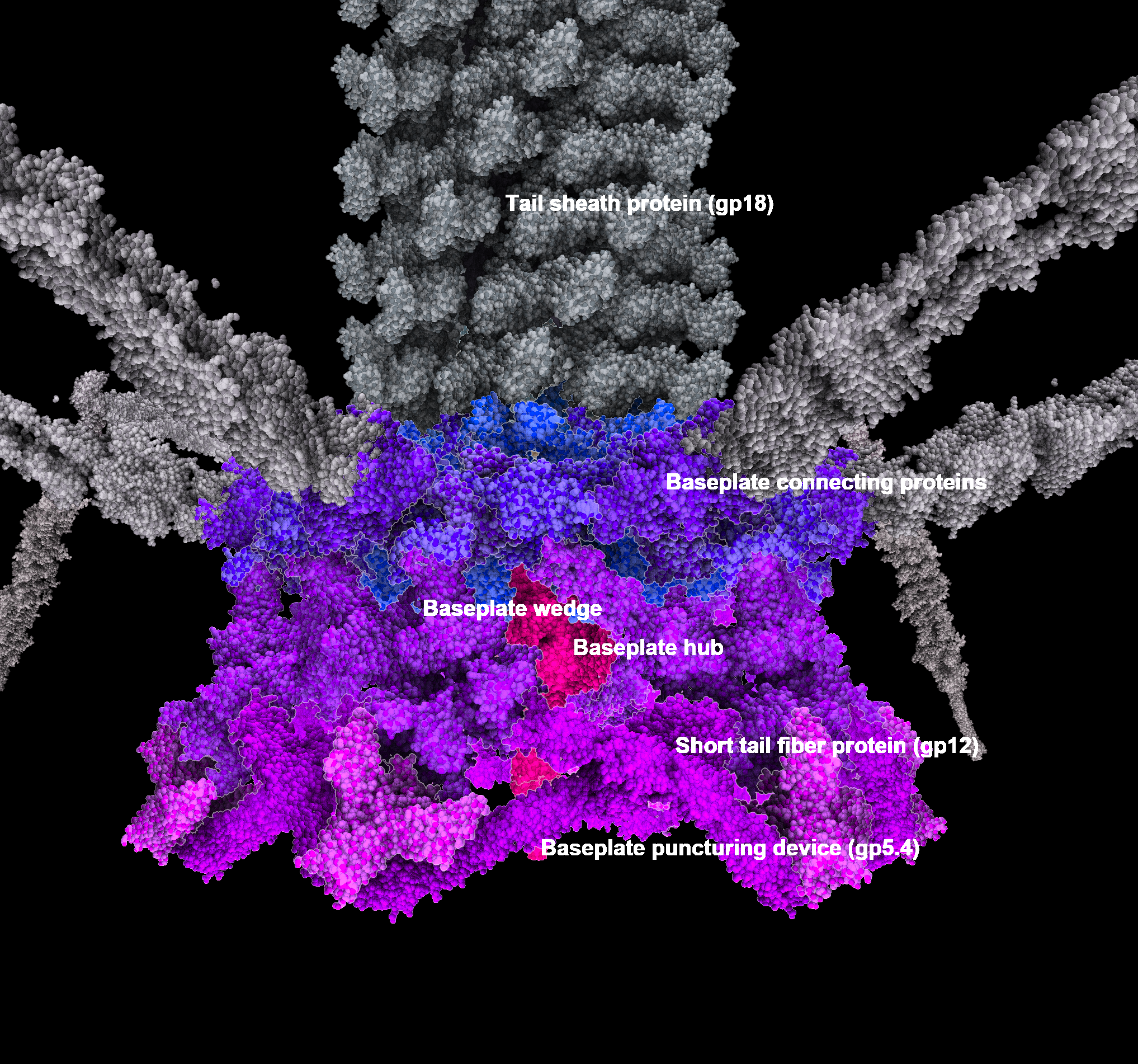}
  }
  \subfigure[SARS-CoV-2 – initial view]{
  \centering
    \includegraphics[width=\optim\linewidth, height=\optim\linewidth]{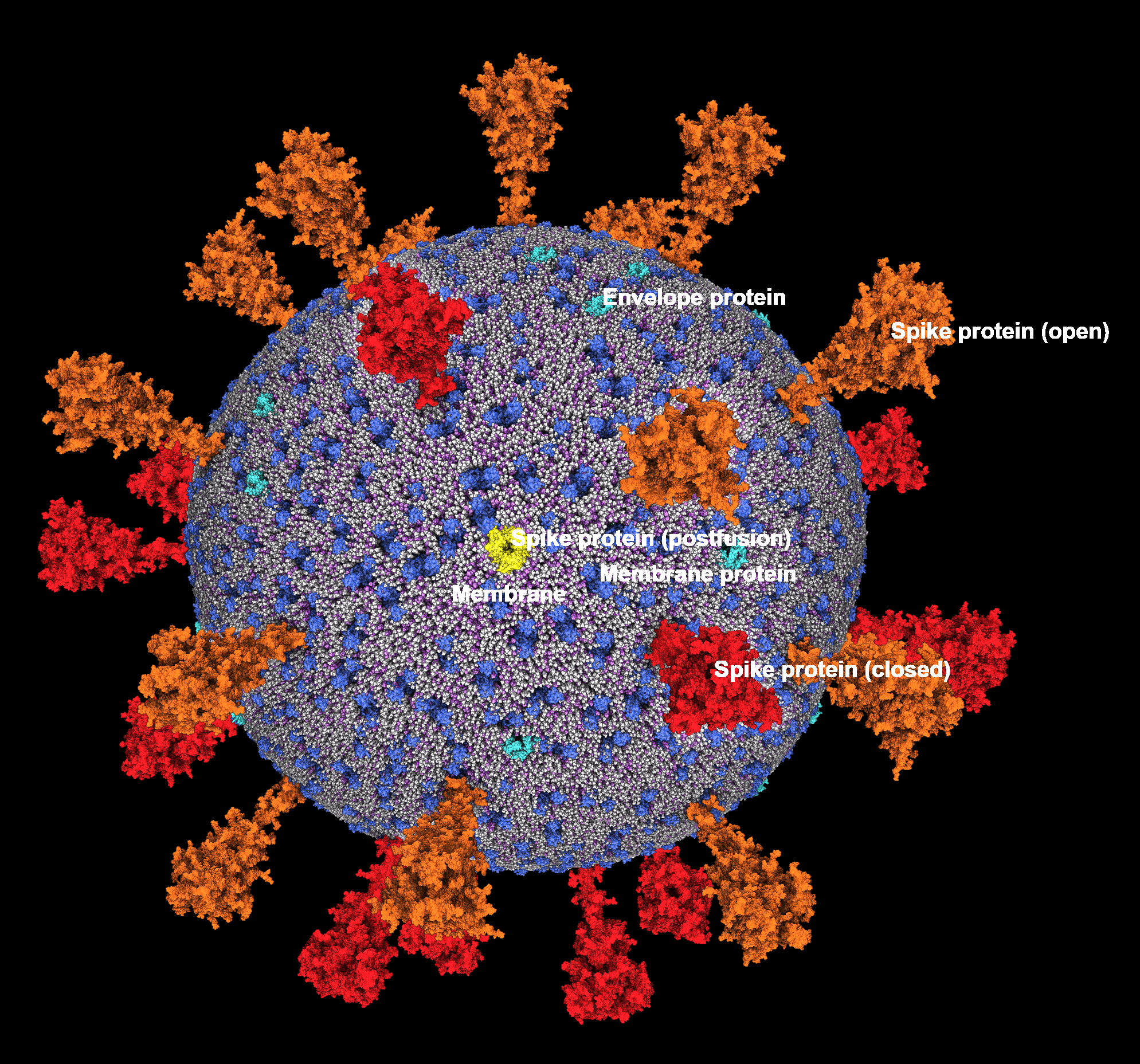}
  }
  \subfigure[Show me the interior objects of the viral lumen.]{
  \centering
    \includegraphics[width=\optim\linewidth, height=\optim\linewidth]{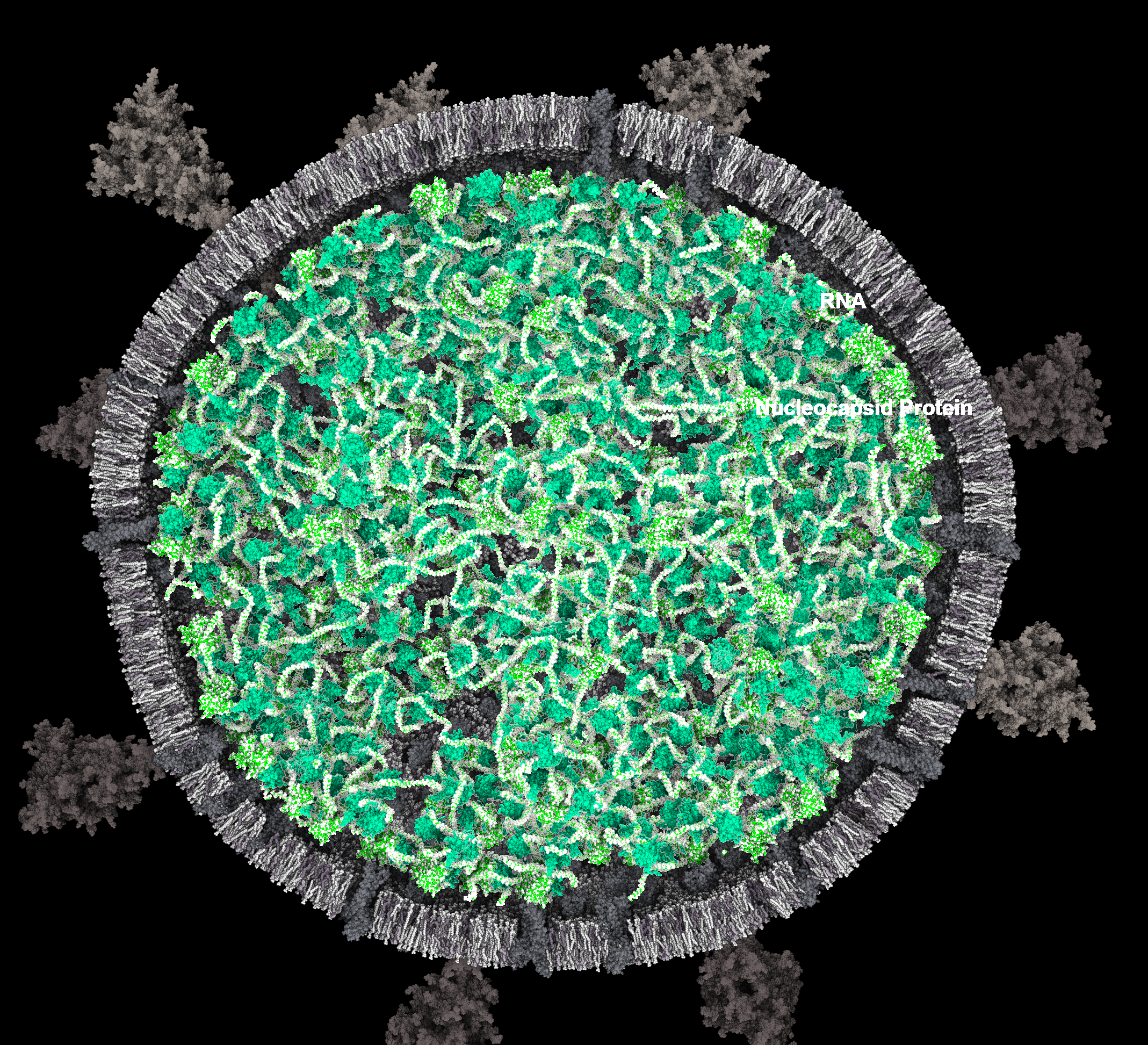}
  }
  \subfigure[Show me the RNA.]{
  \centering
    \includegraphics[width=\optim\linewidth, height=\optim\linewidth]{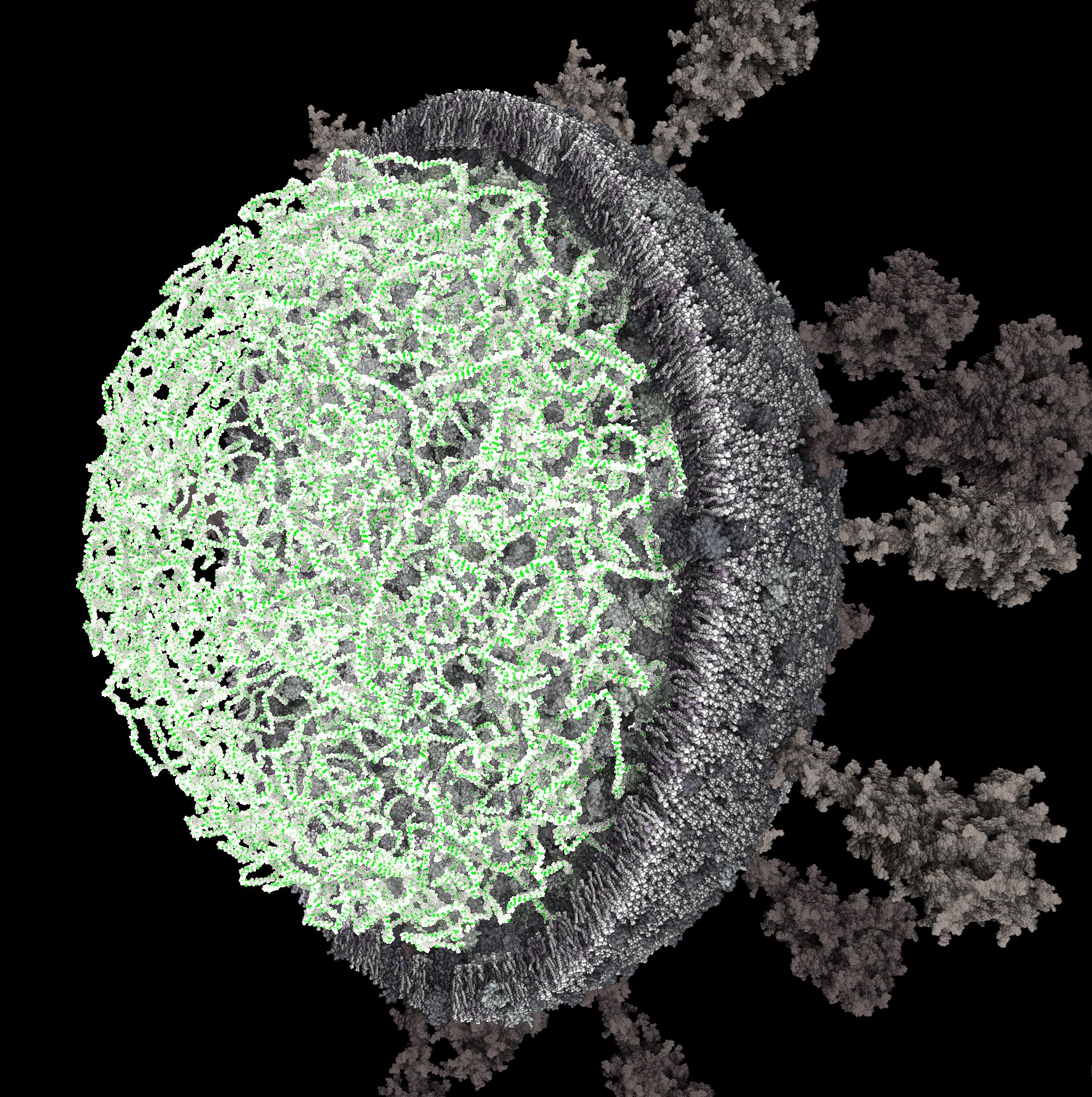}
  }

  \subfigure[HIV – initial view]{
  \centering
    \includegraphics[width=\optim\linewidth, height=\optim\linewidth]{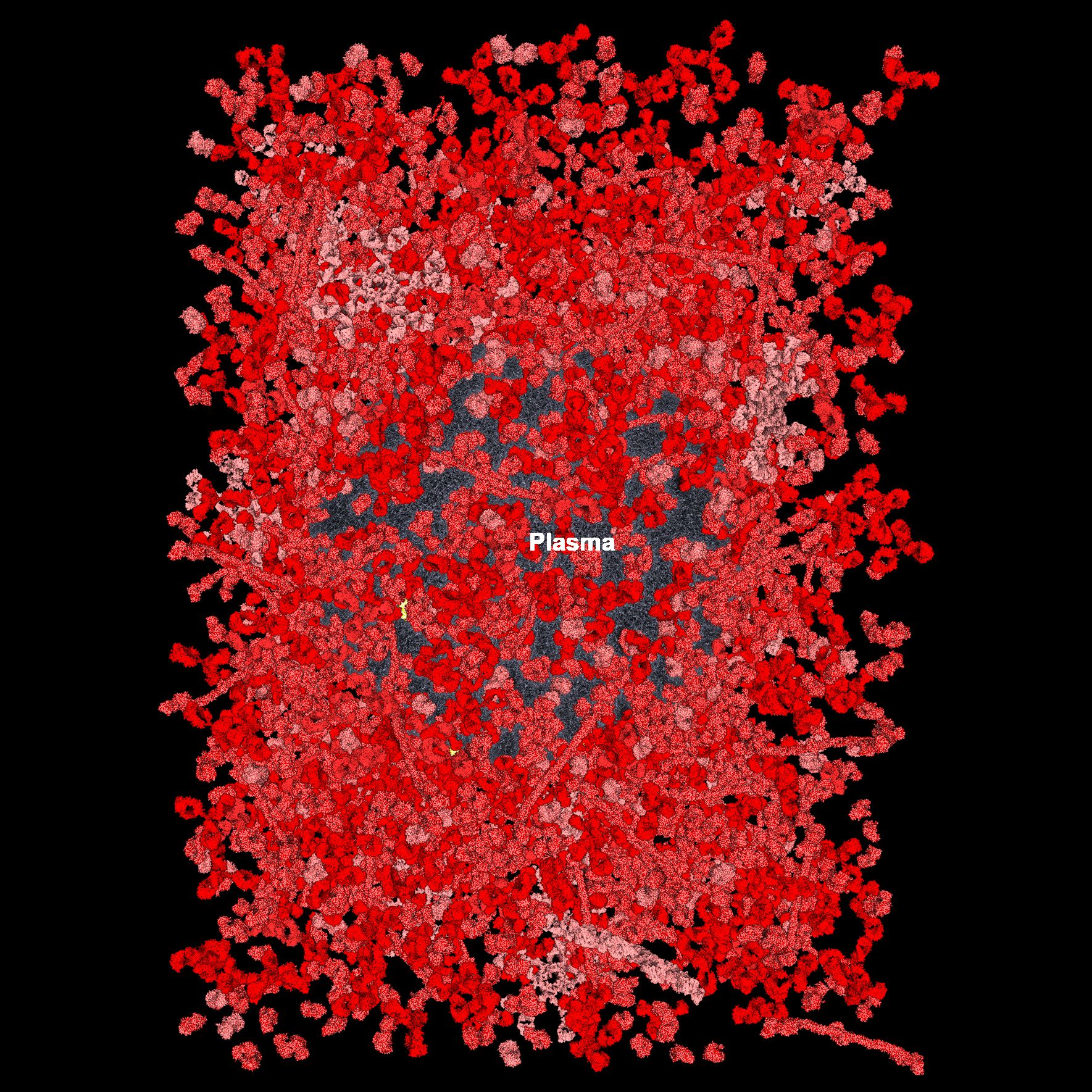}
  }
  \subfigure[Show me how the inside looks.]{
  \centering
    \includegraphics[width=\optim\linewidth, height=\optim\linewidth]{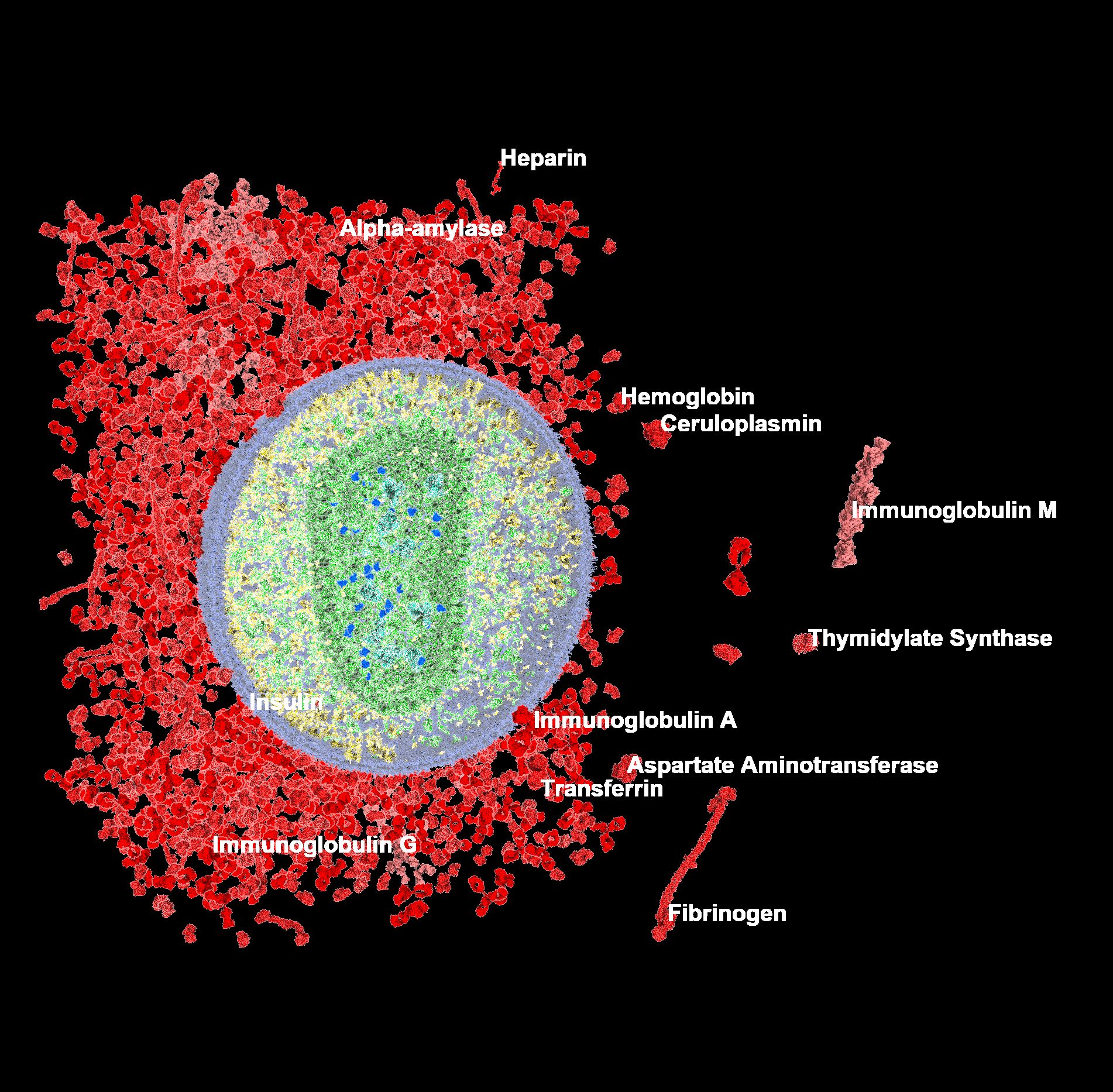}
  }
  \subfigure[Guide me to the matrix protein. ]{
  \centering
    \includegraphics[width=\optim\linewidth, height=\optim\linewidth]{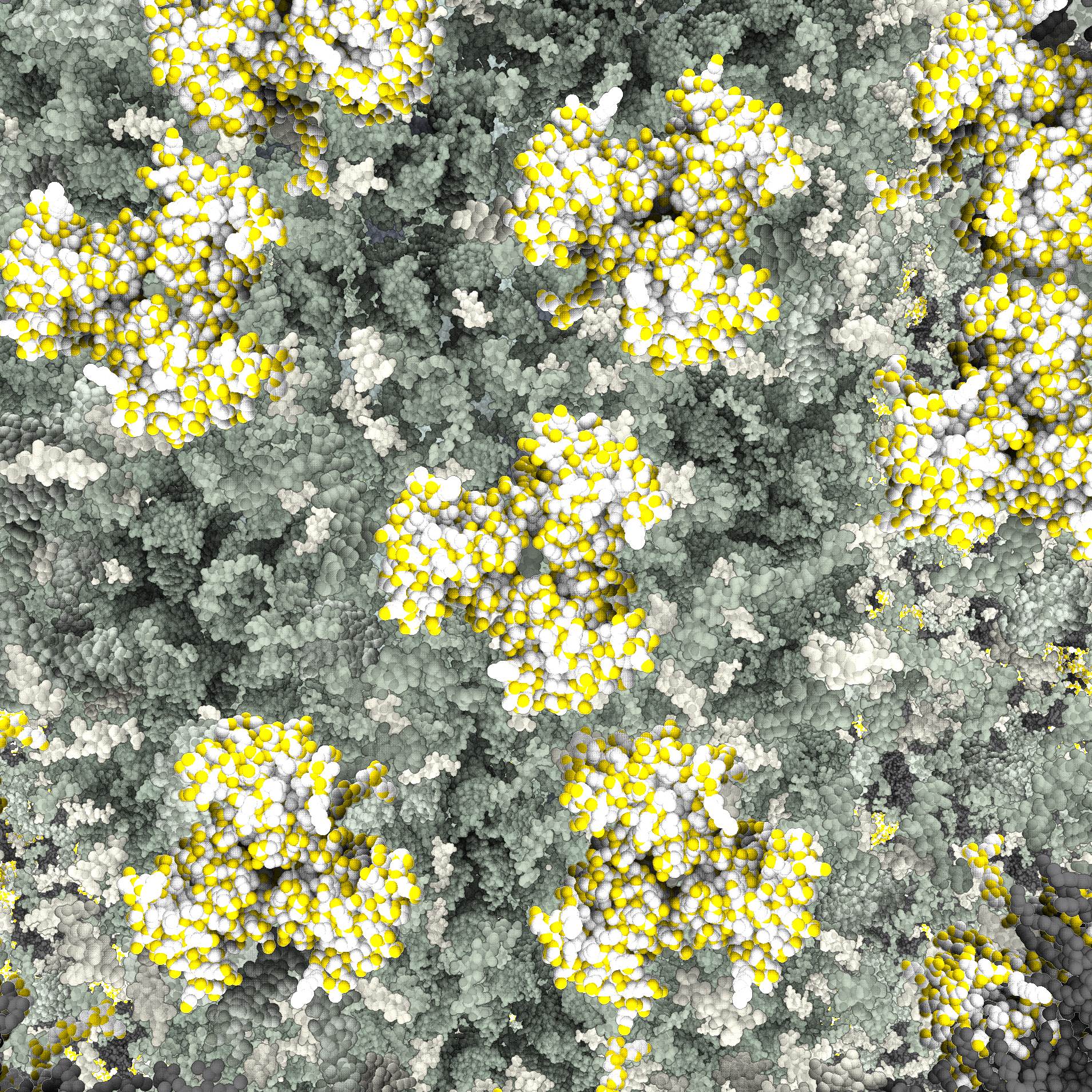}
  }
  \caption{
\textbf{The initial view of three molecular models and demonstration of different commands.} For each model, we demonstrate a multi-turn conversation scenario including either Explorer, Pilot, or cutting actions. 
}
  \label{fig:models}
\end{figure}

We have implemented a prototypical application for molecular data demonstration based on the original Molecumentary~\cite{kouril2021molecumentary} implementation, which was built on the Marion library~\cite{marion} using the Qt framework. For the rendering pipeline, we employ CellVIEW's~\cite{le2015cellview} impostor approach. The performance is measured at full-HD resolution by an NVIDIA RTX 4090 graphics card. 

\subsection{Large Language Model Selection}\label{largelanguagemodelselection}
We conduct a comprehensive comparative analysis involving the GPT series, along with publicly available large language models such as LLaMa-30b-instruct\cite{LLaMa-30b-instruct} and LLaMa-2-70b-instruct-v2\cite{Llama-2-70b-instruct-v2} from the Hugging Face Leaderboard\footnote{\url{https://huggingface.co/spaces/HuggingFaceH4/open_llm_leaderboard}}, as well as smaller LLMs like Alpaca-loRA-7b\cite{alpaca-lora}. Our aim is to identify the optimal model for achieving high-accuracy completion of our designated tasks. Specifically, the LLama-30b-instruct model, comprising 30 billion parameters, is fine-tuned on publicly available question-answering datasets. Similarly, the LLaMa-2-70b-instruct-v2 model, encompassing 70 billion parameters, undergoes fine-tuning on publicly available datasets, demonstrating leading performance on the Hugging Face OpenLLM Leaderboard. Another notable model, Alpaca-lora-7b, incorporates 7 billion parameters and is fine-tuned using the Stanford Alpaca\cite{alpaca} instruction-following data.

Our evaluation encompasses various tasks, including Pilot, Explorer, and Encyclopedia tasks. We directly compare the output to the correct instruction for the instruction extraction task. Conversely, we assess accuracy and consistency in the content generation task. We provide illustrative examples of diverse tasks in supplementary material section 3.

Based on our comparison, small LLMs like Alpca-lora-7b cannot correctly complete our Explorer and Pilot bots' instructions extraction tasks and cannot generate accurate content. Open-sourced LLM with top performance, such as LLaMa-2-70b-instruct-v2, could extract instructions correctly but could not generate high-quality content based on our prompt. Only GPT-4 could accurately extract the instructions and generate high-quality content matching our requirements. We adapt the OpenAI API\footnote{\url{https://platform.openai.com/docs/models}}  to build all of our \textit{pack-of-bots}. We fine-tune the GPT-3.5-turbo version for the \textit{Manager} bot by our own dataset. The dataset contains 473 input-label pairs, covering five classes of our classification task. We split 80\% of the data as the train set and 20\% as the validation set.  For the other bots, we prompt GPT-4. We design the universal prompt for all bots except the \textit{Encyclopedia} bot. For each different model, we add the model description and node names into the \textit{Encyclopedia} bot prompt to improve the quality of the text. 

\subsection{Latency Reduction} \label{reduce latency}

As the model parameters increase, the content generation time of large language models becomes notably extended. While \autoref{largelanguagemodelselection} establishes that larger models yield enhanced performance, the concern arises that users may encounter undesirable delays in receiving responses. In this section, we present two strategies aimed at curtailing the overall query processing response time (algorithm \autoref{alg:queryprocessing}).

As elucidated in algorithm \autoref{alg:queryprocessing}, the engagement of other bots in processing queries follows the acquisition of results from the manager bot. In practical implementation, ensuring the consistent querying time of the manager bot can be challenging, particularly if latencies arise. To mitigate the total query time and enhance stability, we opt for parallelizing the entire query process, allowing all bots to post their queries simultaneously.

Another pertinent issue identified in our implementation is the Encyclopedia bot. A single query often prompts the LLM to generate extensive content, consequently leading to substantial processing time. For instance, GPT-4 requires up to 20 seconds for a solitary query. Yet, inundating users with lengthy responses for each query may prove overwhelming. Despite this, our intent remains to deliver comprehensive answers when users exhibit interest in specific topics. To streamline querying times and enhance user experience, we propose partitioning the Encyclopedia bot's processing into two distinct segments: a concise portion and a detailed segment. Both these segments also become part of the parallelization process. Initially, we furnish a succinct response and inquire whether the user desires further elaboration. Should the response be affirmative, we subsequently provide the detailed segment to satisfy the user's curiosity.

Following optimization, we have achieved varying degrees of acceleration in the processing time for each bot. The delay duration for the Explorer and Pilot bots has been reduced to approximately one and two seconds, respectively. Moreover, the latency for the Encyclopedia bot has been notably diminished, now operating at a satisfied 3-4 seconds. As a result, users are experiencing minimal delays when utilizing the system, fostering a more seamless and efficient user experience.
\subsection{User Experience}
In addition to selecting an appropriate language model and reducing latency to enhance the user experience, we have implemented several additional strategies related to speech, labeling, and narration to elevate user engagement and satisfaction further. These strategies directly contribute to addressing design requirement 5 (guiding) and promoting a more intuitive exploration process for users.

We utilize the available Qt component\footnote{\url{https://doc.qt.io/qt-5/qaudiorecorder.html}} to facilitate voice recording. To convert the speech signal to text, we employ the OpenAI Whisper API\footnote{\url{https://platform.openai.com/docs/guides/speech-to-text}}. The Whisper API supports giving a prompt to improve the quality of the transcripts. Therefore, we provide multiple biology jargon about our model as the prompt and get better performance recognizing the biology-related word. For text-to-speech functionality, we utilize  Google's Cloud text-to-speech API\footnote{\url{https://cloud.google.com/text-to-speech/docs/reference/rest/}}, which provides high-fidelity speech output that approaches human quality and offers a wide selection of voices to match varying demands.

With only a molecular model displayed in the scene, users might not know what to ask and what to explore. To guide the user to explore our model, we take several strategies: first is labeling. We adapt HyperLabels~\cite{kouvril2020hyperlabels} as the labeling approach. The method could place the labels of the ingredients in the proper position based on the ingredient depth, ingredient position, and scene tree structure. This textual information ensures the user gets a sketchy understanding of the model and then uses our VOICE method for further exploration.

The second strategy to guide the user is the guidance narration. Without any help, users might not know how to interact with the VOICE system. We design the introduction narration and show it when the application starts, introducing the basic information of the current model and providing basic operations. We also design a help function if the user needs any further help, introducing more details about how to use the application. Between each question-answering round, we design some transition narration to avoid blunt conversion. All the designed narration can be found in supplementary material section 4.
\subsection{Data} \label{Data}
We evaluate our \methodname{} framework on three molecular models. To demonstrate our framework's capabilities comprehensively, we record real-time videos of all datasets at full-HD resolution. These demonstrations are included in the supplementary demo videos. 

First, we use a \textbf{Bacteriophage T4} model (\autoref{fig:models} (a)-(c)) provided by KAUST~\cite{NanovisT4Model}. This structural model comprises approximately 700 distinct elements, consisting of roughly 6.2 million atoms, not including lipids. It is worth noting that the double-stranded DNA is not present in the model since the authors did not create the DNA file.  The scene tree contains 39 nodes overall, 31 leaf nodes, and eight higher-level nodes.

Next, we utilize a \textbf{SARS-CoV-2} model (\autoref{fig:models} (d)-(f)) also provided by KAUST~\cite{nguyen2021modeling}. This structural model comprises approximately 200,000 elements and includes lipids totaling around 14 million atoms. The authors also modeled the single-stranded RNA. For the scene tree file, we create 16 nodes overall, five protein types, one phosphate and sugar type, two lipid types, four nucleotide types, and three higher-level nodes.

Finally, we use an \textbf{HIV in blood plasma} model (\autoref{fig:models} (g)-(i)) from Scripps Research. This structural model comprises approximately 18,500 protein instances and 200,000 lipids. We build a scene tree with 42 nodes—35 protein types, two lipid types, and five higher-level nodes.

\section{Expert Evaluation}
\label{sec:user_study}

Beyond validating that \appname\ can render and interact with specific datasets, it is important to validate that the prototype also answers the specific needs identified by our expert educators, particularly concerning the design requirements. As such, we detail our expert evaluation below.

\subsection{Evaluation process}

To evaluate our system, we asked the three educators we interviewed in \autoref{sec:design_requirements} to participate in an evaluation of our prototype. We decided to ask the same three experts for two main reasons. First, their extensive experience and knowledge of public dissemination, pedagogical research, and systems in public science centers are essential and rare backgrounds to be able to assess our prototype with real use cases in mind. Second, we thought that this way, they may be able to reflect on some of the design requirements they previously helped highlight and provide us with extensive and insightful feedback. 

Our evaluation consisted of quickly presenting \appname\ to the three experts (without revealing too much about what it can or cannot do) and asking them to try using it on two of our datasets (namely the T4 and SARS-CoV-2). We told them they could try \appname\ for as long as they desired and asked them to reflect on its functions and answer aloud so that the facilitator in the room could take notes of their reflections. Each evaluation lasted between 60 and 90 minutes and was concluded by semi-structured interviews. As is usual with semi-structured interviews, the facilitator made use of these questions only to elicit specific feedback and comments if they were not mentioned earlier and adapted to the conversation that the expert and they were having. We chose this semi-structured interview methodology because of its potential to highlight interest, ideas, and feedback of new technologies \cite{reinschluessel2022virtual,coulentianos2020stakeholder}.

\subsection{Results}

From this set of semi-structured interviews, we can extract specific findings that are particularly interesting to evaluate the usefulness and usability of \appname\ as well as helpful or missing features or content that can help us draw the landscape of our future work. We develop on these below.

\textbf{Latency}. All three experts highlighted that the current latency of the system was not a problem at all for them and that they could easily envision using the system in both of the scenarios they considered, namely the \emph{guided} and \emph{unguided} ones. In the \emph{guided} scenario, the narrator would carefully use trigger when he talks to \appname\ and then let the system process and dynamically answer to show and navigate to the needed scene while the narrator continues to deliver explanations. As such, the latency would not be an issue. In particular, E1 mentioned that ``[Narrators] would have to learn and adjust [their] speech at the beginning, but it would eventually work. It's really easy to work around.'' In the \emph{unguided} scenario, the fact that \appname\ usually indicates when it needs to process a more complicated query and ask for the users to wait ensures that latency is not an issue. All participants mentioned that the trigger mechanism makes the system easy to use and realistic in the environment of a public space in addition to fulfilling some of the requirements and thoughts they had in \autoref{sec:design_requirements}. 

\textbf{Usability}. Our three experts highlighted that the hints given by the \appname\ on what can be explored, while useful in the beginning, are too repetitive, too long, and should be more context-aware (e.g., only propose hints when the user seems to be lost). On this topic, E3 stated: ``How do I know that I explored enough? Maybe it should steer things more a bit in some cases when it sees that you are not exploring things.'' One of our experts (E2) highlighted that the step-by-step navigation should propose to go more in-depth within the structure of the virus. 
In this regard also, while this step-by-step navigation that \appname\ fosters is particularly useful with a knowledgeable narrator (i.e., the \emph{guided} scenario in which \appname\ can act as a co-pilot assisting the narrator), other possibilities should be presented for interactions with lay audiences such as the possibility to answer a query such as  ``Give me a short story of this virus'' (E3) or directly propose ``a short animation to present the virus'' (E2). One main issue for lay audiences also lies in the complex vocabulary needed to activate some commands with \appname.  
One expert (E2) also mentioned that there could be some frustration induced by the fact that sometimes when the system does not quite understand what you want and starts a long answer, you will have to wait until it is done talking to try again. This is interesting feedback since we did not tell experts that they would have to wait until the system is done talking to send in a new command. This may further demonstrate users' abilities to anthropomorphize ChatGPT \cite{cheng2022human,jacobs2023brief,ma2023users} and is definitely an interesting finding that should be further studied in the specific context of educational assistants like \appname. While trying to send a command during \appname's speech might currently make the prototype wrongly listen to itself, it would be easy to modify our implementation such that a click on the ``record audio'' trigger stops \appname's narration. However, while such verbal patterns and interactions might mimic the way humans can interrupt each other during conversations, discussions and conversations in science centers usually follow different codes and patterns and are currently an active research topic \cite{krange2020peers,schiele2021science,shaby2019examination,shaby2019engagement}. However, our findings prompt us to believe that investigating the benefits and drawbacks of different conversation models between people and verbal chatbots in public centers should be carefully studied and represent a significant avenue for future work. As such, \appname\ also shows the potential to enrich the educational literature.

\textbf{Content}. According to our three experts, the answers given by \appname\ about biomolecular concepts were accurate. One of our experts (E1) was even surprised about the accuracy and knowledge displayed by our prototype. Our three experts highlighted that the content was accurate and complete enough to understand the structure, but additional context knowledge might be wanted. For instance, it would be interesting to be able to have a more accurate breakdown of the different sequences of how a virus infects a cell. The experts all highlighted how difficult it is to understand and convey processes around infection or immune response. This is not something we set out to achieve through \appname\, but it seems that a careful co-design process with our educators might help achieve this goal. 
In addition, all experts expressed that they understood that the visualization could not embed all elements around viral infection or immune responses or location within the model visualized. Yet, they expressed that the system would be enriched by proposing pre-computed still images or short animations that would complement the visual presentation of the virus' structure with other contextual visualizations. As such, in context, \appname\ could show a short animation of how the T4 attaches to a host cell, for instance, and transmits its genetic materials while simultaneously providing verbal explanations and contextual focuses on the visualized data. Such animations would be focus-dependent and have to be pre-computed and loaded into the program but could be made with any software and do not represent a major implementation challenge. 
Finally, our experts highlighted that the vocabulary might need to be adapted in the case of \emph{unguided} exploration such that it can be understood by all visitors, regardless of their knowledge.

\textbf{Features and commands}. We asked our experts what they particularly appreciated and which ones they believed were missing for the system to be deployed in an informal educational context. 
All three indicated that the verbal interface was exciting, fun, or entertaining, and visitors will likely enjoy using it. Of course, the novelty effect may be responsible for this appreciation (see, e.g., \cite{besancon:hal-01436206}), yet this property of our system may very well help retain visitors at exhibits longer and, therefore, potentially increase their educational value \cite{seale2023education,Xu:2022}. However, as this potential has been envisioned and eventually mitigated for other technologies in public learning settings (e.g., VR \cite{shahab2023virtual}), this also represents an important avenue for future investigations.
The three experts also indicated that they appreciated the visioverbal synchronized explanations given by the system, although they wished that some of it could have gone more in-depth within the hierarchies of viruses' structures.
On the side of missing features or commands, our experts highlighted some basic and more complicated features that may be good to implement:
\begin{itemize}
    \item[F1] A command to automatically, infinitely, and slowly spin the model to improve depth perception and 3D understanding of the structure. 
    
    \item[F2] The possibility to highlight or filter out specific structures (with color/transparency or a selection shape) and combining this possibility with the narration that the system proposes such that visitors do not have to focus on the visual representation to figure out what \appname\ is talking about.

    \item[F3] A command that would allow users to change the representation from Van der Waals spheres to the ribbon (for instance) fully or partially, allowing them to highlight specific structures or regions and simplify understanding through hybrid representations. 

    \item[F4] The possibility to directly manipulate the cutting plane throughout the structure (as an animation or as successive commands to cut deeper). 
\end{itemize}

Overall, our three experts were pleasantly surprised about the prototype and excited about the possibilities it offered both for \emph{guided} exploration (in which the system would just be a co-pilot and not talk) and for \emph{unguided} exploration. E1 even went as far as to state that the system could be used as a co-teacher, thus envisioning a new use case for the system. The educators all highlighted that the system was usable, that latency was not an issue, and that the content was already accurate and helpful while requesting additional features and content before deployment, some of which we will discuss further below.  

\section{Discussion} \label{sec:discussion}

Our VOICE approach is motivated by the desire to make scientific discoveries generally accessible to a diverse audience, especially users without specific domain knowledge, for example, non-experts in structural biology. Based on the expert evaluation we conducted with three educators, we discuss how our design requirements have been fulfilled or how they could be better addressed.  

Our expert evaluation highlighted that our prototype completely fulfills design requirements 1 to 3, namely that the system should act as a pilot interactively steering the visualization system, be able to operate the view and visualization widget manipulation, and have contextual awareness of the presented data. Implementing F1, as suggested by our experts, increases the fulfillment of design requirement 1 (steering) and has already been implemented in our revised version of \appname. Similarly, implementing F4 might improve the fulfillment of design requirement 2 (view) and should not be particularly too complicated to implement. However, past work suggests that direct interaction paradigms might better handle such precise manipulation \cite{Keefe:2013:RSV}, thus rendering voice a suboptimal choice when implementing F4. We are consequently considering providing a multi-modal interaction mechanism \cite{Besancon:2021:STAR} that would combine touch/mouse and voice interaction and allow the conduct of precise cutting plane manipulation through the former. This input combination has rarely been studied in the literature \cite{Besancon:2021:STAR} and only a handful of examples can be found for visualization purposes \cite{Lubos:2014:TC,Tsang:2002:BOOM}, and none of these examples really used voice input as a way to manipulate data as our prototype does. 

Design requirements 4 and 5, namely, the system should be able to change representation modes and metaphors, be intuitive, and offer initial guidance for exploration, are only partially fulfilled and would directly benefit from implementing F2 and F3. These are, in turn, likely to improve the understanding of complex structures since different levels of visual literacy and pre-knowledge may require different representations \cite{coan2020teaching,jimenez2019teaching,schonborn2010bridging}. However, these implementations are not trivial and would require more work on our part and a better understanding of how educators use different representations and highlighting features so that we could co-design our system to support their exact use cases.

As we decided to postpone addressing design requirement 6 (the system should adapt to its audience and their knowledge), it is not surprising that the educators highlighted that for an \emph{unguided} exploration, the vocabulary might be too complicated. However, this is a challenge that all exhibits, regardless of their focus or technology, face but this may be where LLMs and systems such as ours can actually assist in finding a dynamic solution. Should we manage to achieve this, we could ensure that visitors could be delivered with knowledge-adapted vocabulary that would scaffold their learning. 

While we also decided to postpone addressing design requirement 7 (the prototype should also be analyzed and adapted to group dynamics) because of its technological and theoretical complexity, the issue did not arise from our evaluation. Nonetheless, group and conversational dynamics in public centers are essential components of learning \cite{Sobel:2021:RBPC,shaby2019examination}, be it between educators and visitors \cite{massarani2022role,shaby2019examination} or within groups of visitors such as students/pupils or families \cite{blud1990social,joy2021understanding,Sobel:2021:RBPC}. However, allowing a system like \appname\ to dynamically identify and recognize different voices while also analyzing the content of each spoken sentence such that it could understand the relational dynamics of the conversation is particularly complicated. Voice recognition is already complex even within a relatively quiet environment, even more so in a public space, which is thus bound to be even more difficult with the current technology. Further, text analysis of conversation might also prove to be its own challenge as it is still a growing area of research.

Our evaluation of \appname\ has obvious limitations. First of all, we only recruited a small number of experts to evaluate the system. However, access to expert users is always a complicated issue with HCI and visualization research, and as such, such low sample sizes are not uncommon \cite{Caine:2016:LSS,Isenberg:2013,koeman2018participants} and perfectly allow for assessment benefits and limitations of specific prototypes \cite{besancon:hal-02381513,Besancon:2017:HTT,pooryousef2023working}).
Another obvious limitation lies in the fact that we did not evaluate our prototype with lay audiences or in the wild. However, such an evaluation would involve already having a system that would satisfy the team of expert educators that we interviewed and, therefore, falls outside this paper's scope. However, these evaluations would need to be conducted to ensure that the system uses appropriate vocabulary, meaningful visioverbal communication, and provides an enjoyable experience for museum visitors and public spaces. Further evaluation should also include assessing learning outcomes from a system such as ours compared to a more traditional museum exhibit.

\section{Conclusion and Future Work}

In the past, conversational visualization systems were mostly applied to tabular data \cite{FacilitatingConvInteraction,ADVISor,NLtoVISbyNMT,NL2VIS,InChorus,wang2022towards,maddigan2023chat2vis}. To the best of our knowledge, we are the first to connect large language models' conversational capabilities with 3D visualization for science communication through this work. We support users in starting a multi-turn conversation to explore and interact with 3D molecular models. By utilizing our pack-of-bots strategies, the described VOICE method can parse arbitrary user input and provide highly accurate responses from specific bots. We also propose a new interactive text-to-visualization approach to match generated text with high fidelity to our visualization output. By applying the VOICE method, we aim to provide a highly flexible, informative, and educational tool for the general public to explore complex and dense visual representations. We exemplify this by applying our methods to complex molecular models. 

A natural next step for future work would involve instruction extraction and task assignment in the large language model training stage. This would allow for one pre-trained large language model to handle all commands. Although this would limit the current flexibility of our approach and make it more difficult to add new types of commands, it would likely also result in better performance on current tasks.

Currently, the dialogue system and visualization system are separated. A further step would be to embed the visual information into a feature and feed it into the dialogue system, allowing the dialogue system's decision-making to involve the visual representation, such as the current scale information, position and color of objects, and angle of view. This would lead to a natural convergence of the visual and verbal language.

A  priority for future research is to incorporate dynamic models into our visualization system and develop an animation auto-generation algorithm. As dynamics and time dependence are critical components of understanding complex processes, doing so would significantly enhance the effectiveness of knowledge dissemination and make our system more appealing to a wider audience.

\section*{Acknowledgments}
{
The authors would like to thank David Kou\v{r}il for providing the Molecumentary framework, as well as Nanographics GmbH for providing the marion framework. 

The models (T4, SARS-CoV2, HIV) were kindly made available by Aeliya Syed, John Winfer, Leon Thistle, Paul Ekers, Deng Luo, Ond\v{r}ej Strnad, Ngan Nguyen, David Goodsell, Martina Maritan, Ludovic Autin, Arthur Olson, and Graham Johnson. The work was supported by the Knut and Alice Wallenberg Foundation (grant KAW 2019\discretionary{.}{}{.}0024).

}
\bibliographystyle{plain}

\bibliography{bibliograph}

\begin{thebibliography}{10}

\bibitem{Besancon:2017:HTT}
Lonni Besan{\c c}on, Paul Issartel, Mehdi Ammi, and Tobias Isenberg.
\newblock {Hybrid Tactile/Tangible Interaction for 3D Data Exploration}.
\newblock {\em {IEEE Transactions on Visualization and Computer Graphics}},
  23(1):881--890, 2017.

\bibitem{besancon:hal-01436206}
Lonni Besan{\c c}on, Paul Issartel, Mehdi Ammi, and Tobias Isenberg.
\newblock {Mouse, Tactile, and Tangible Input for 3D Manipulation}.
\newblock In {\em {Proc.\ CHI}}, pages 4727--4740, Denver, United States, May
  2017.

\bibitem{besanccon2022exploring}
Lonni Besan{\c{c}}on, Konrad Sch{\"o}nborn, Erik Sund{\'e}n, He~Yin, Samuel
  Rising, Peter Westerdahl, Patric Ljung, Josef Widestr{\"o}m, Charles Hansen,
  and Anders Ynnerman.
\newblock Exploring and explaining climate change: Exploranation as a
  visualization pedagogy for societal action.
\newblock In {\em VIS4GOOD, a workshop on Visualization for Social Good, held
  as part of IEEE VIS 2022}, 2022.

\bibitem{besancon:hal-02381513}
Lonni Besan{\c c}on, Amir Semmo, David~J. Biau, Bruno Frachet, Virginie Pineau,
  El~Hadi Sariali, Marc Soubeyrand, Rabah Taouachi, Tobias Isenberg, and Pierre
  Dragicevic.
\newblock {Reducing Affective Responses to Surgical Images and Videos Through
  Stylization}.
\newblock {\em {Computer Graphics Forum}}, 39(1):462--483, January 2020.

\bibitem{Besancon:2021:STAR}
Lonni Besan{\c c}on, Anders Ynnerman, Daniel~F Keefe, Lingyun Yu, and Tobias
  Isenberg.
\newblock {The State of the Art of Spatial Interfaces for 3D Visualization}.
\newblock {\em {Computer Graphics Forum}}, 40(1):293--326, February 2021.

\bibitem{blud1990social}
Linda~M Blud.
\newblock Social interaction and learning among family groups visiting a
  museum.
\newblock {\em Museum Management and Curatorship}, 9(1):43--51, 2009.

\bibitem{bock2020openspace}
Alexander Bock, Emil Axelsson, Jonathas Costa, Gene Payne, Micah Acinapura,
  Vivian Trakinski, Carter Emmart, Cláudio Silva, Charles Hansen, and Anders
  Ynnerman.
\newblock Openspace: A system for astrographics.
\newblock {\em IEEE Transactions on Visualization and Computer Graphics},
  26(1):633--642, 2020.

\bibitem{bock2018openspace}
Alexander Bock, Emil Axelsson, Carter Emmart, Masha Kuznetsova, Charles Hansen,
  and Anders Ynnerman.
\newblock Openspace: Changing the narrative of public dissemination in
  astronomical visualization from what to how.
\newblock {\em IEEE computer graphics and applications}, 38(3):44--57, 2018.

\bibitem{Bottinger2020}
Michael B{\"o}ttinger, Helen-Nicole Kostis, Maria Velez-Rojas, Penny Rheingans,
  and Anders Ynnerman.
\newblock {\em Reflections on Visualization for Broad Audiences}, pages
  297--305.
\newblock Springer International Publishing, Cham, 2020.

\bibitem{Brossier:2023:MOL}
Mathis Brossier, Robin Sk{\aa}nberg, Lonni Besan{\c c}on, Mathieu Linares,
  Tobias Isenberg, Anders Ynnerman, and Alexander Bock.
\newblock {Moliverse: Contextually embedding the microcosm into the universe}.
\newblock {\em {Computers and Graphics}}, 112:22--30, May 2023.

\bibitem{Caine:2016:LSS}
Kelly Caine.
\newblock Local standards for sample size at chi.
\newblock In {\em Proc.\ CHI}, CHI '16, pages 981--992, New York, NY, USA,
  2016. ACM.

\bibitem{cheng2022human}
Xusen Cheng, Xiaoping Zhang, Jason Cohen, and Jian Mou.
\newblock Human vs. ai: Understanding the impact of anthropomorphism on
  consumer response to chatbots from the perspective of trust and relationship
  norms.
\newblock {\em Information Processing \& Management}, 59(3):102940, 2022.

\bibitem{coan2020teaching}
Heather~A Coan, Geoff Goehle, and Robert~T Youker.
\newblock Teaching biochemistry and molecular biology with virtual
  reality—lesson creation and student response.
\newblock {\em Journal of Teaching and Learning}, 14(1):71--92, 2020.

\bibitem{coulentianos2020stakeholder}
Marianna~J Coulentianos, Ilka Rodriguez-Calero, Shanna~R Daly, and Kathleen~H
  Sienko.
\newblock Stakeholder engagement with prototypes during front-end medical
  device design: Who is engaged with what prototype?
\newblock In {\em Frontiers in Biomedical Devices}, volume 83549, page
  V001T08A001. American Society of Mechanical Engineers, 2020.

\bibitem{hoest2020nano}
Gunnar Höst, Karljohan Palmerius, and Konrad Schönborn.
\newblock Nano for the public: An exploranation perspective.
\newblock {\em IEEE Computer Graphics and Applications}, 40(2):32--42, 2020.

\bibitem{Isenberg:2013}
T.~{Isenberg}, P.~{Isenberg}, J.~{Chen}, M.~{Sedlmair}, and T.~{Möller}.
\newblock A systematic review on the practice of evaluating visualization.
\newblock {\em IEEE Transactions on Visualization and Computer Graphics},
  19(12):2818--2827, Dec 2013.

\bibitem{jacobs2023brief}
Oliver Jacobs, Farid Pazhoohi, and Alan Kingstone.
\newblock Brief exposure increases mind perception to chatgpt and is moderated
  by the individual propensity to anthropomorphize.
\newblock 2023.

\bibitem{jimenez2019teaching}
Zulma~A Jim{\'e}nez.
\newblock Teaching and learning chemistry via augmented and immersive virtual
  reality.
\newblock In {\em Technology Integration in Chemistry Education and Research
  (TICER)}, pages 31--52. ACS Publications, 2019.

\bibitem{joy2021understanding}
Angelina Joy, Fidelia Law, Luke McGuire, Channing Mathews, Adam Hartstone-Rose,
  Mark Winterbottom, Adam Rutland, Grace~E Fields, and Kelly~Lynn Mulvey.
\newblock Understanding parents’ roles in children’s learning and
  engagement in informal science learning sites.
\newblock {\em Frontiers in Psychology}, 12:635839, 2021.

\bibitem{Keefe:2013:RSV}
Daniel~F. Keefe and Tobias Isenberg.
\newblock Reimagining the scientific visualization interaction paradigm.
\newblock {\em IEEE Computer}, 46(5):51--57, May 2013.

\bibitem{koeman2018participants}
Lisa Koeman.
\newblock How many participants do researchers recruit? a look at 678 ux/hci
  studies.
\newblock Online. Last visited 06 January 2019, 2018.

\bibitem{kouvril2020hyperlabels}
David Kou{\v{r}}il, Tobias Isenberg, Barbora Kozl{\'\i}kov{\'a}, Miriah Meyer,
  M~Eduard Gr{\"o}ller, and Ivan Viola.
\newblock Hyperlabels: Browsing of dense and hierarchical molecular 3d models.
\newblock {\em IEEE Transactions on Visualization and Computer Graphics},
  27(8):3493--3504, 2020.

\bibitem{kouril2021molecumentary}
David Kouril, Ondrej Strnad, Peter Mindek, Sarkis Halladjian, Tobias Isenberg,
  Eduard Groeller, and Ivan Viola.
\newblock Molecumentary: Adaptable narrated documentaries using molecular
  visualization.
\newblock {\em IEEE Transactions on Visualization \& Computer Graphics},
  (01):1--1, 2021.

\bibitem{Krange2020PTG}
Ingeborg Krange, Kenneth Silseth, and Palmyre Pierroux.
\newblock Peers, teachers and guides: A study of three conditions for
  scaffolding conceptual learning in science centers.
\newblock {\em Cultural Studies of Science Education}, 15(1):241--263, 2020.

\bibitem{krange2020peers}
Ingeborg Krange, Kenneth Silseth, and Palmyre Pierroux.
\newblock Peers, teachers and guides: A study of three conditions for
  scaffolding conceptual learning in science centers.
\newblock {\em Cultural Studies of Science Education}, 15:241--263, 2020.

\bibitem{le2015cellview}
Mathieu Le~Muzic, Ludovic Autin, Julius Parulek, and Ivan Viola.
\newblock Cellview: a tool for illustrative and multi-scale rendering of large
  biomolecular datasets.
\newblock In {\em Eurographics Workshop on Visual Computing for Biomedicine},
  volume 2015, page~61. NIH Public Access, 2015.

\bibitem{ADVISor}
Can Liu, Yun Han, Ruike Jiang, and Xiaoru Yuan.
\newblock Advisor: Automatic visualization answer for natural-language question
  on tabular data.
\newblock In {\em 2021 IEEE 14th Pacific Visualization Symposium (PacificVis)},
  pages 11--20, 2021.

\bibitem{Lubos:2014:TC}
P.~Lubos, R.~Beimler, M.~Lammers, and F.~Steinicke.
\newblock Touching the cloud: Bimanual annotation of immersive point clouds.
\newblock In {\em Proc.\ 3DUI}, pages 191--192, Los Alamitos, 2014. IEEE
  Computer Society.

\bibitem{NL2VIS}
Yuyu Luo, Nan Tang, Guoliang Li, Chengliang Chai, Wenbo Li, and Xuedi Qin.
\newblock Synthesizing natural language to visualization (nl2vis) benchmarks
  from nl2sql benchmarks.
\newblock In {\em Proceedings of the 2021 International Conference on
  Management of Data}, SIGMOD '21, page 1235–1247, New York, NY, USA, 2021.
  Association for Computing Machinery.

\bibitem{NLtoVISbyNMT}
Yuyu Luo, Nan Tang, Guoliang Li, Jiawei Tang, Chengliang Chai, and Xuedi Qin.
\newblock Natural language to visualization by neural machine translation.
\newblock {\em IEEE Transactions on Visualization and Computer Graphics},
  28(1):217--226, 2022.

\bibitem{ma2012plankton}
Joyce Ma, Isaac Liao, Kwan-Liu Ma, and Jennifer Frazier.
\newblock Living liquid: Design and evaluation of an exploratory visualization
  tool for museum visitors.
\newblock {\em IEEE Transactions on Visualization and Computer Graphics},
  18(12):2799--2808, 2012.

\bibitem{ma2020plankton}
Joyce Ma, Kwan-Liu Ma, and Jennifer Frazier.
\newblock Decoding a complex visualization in a science museum – an empirical
  study.
\newblock {\em IEEE Transactions on Visualization and Computer Graphics},
  26(1):472--481, 2020.

\bibitem{ma2023users}
Xiaoyue Ma and Yudi Huo.
\newblock Are users willing to embrace chatgpt? exploring the factors on the
  acceptance of chatbots from the perspective of aidua framework.
\newblock {\em Technology in Society}, 75:102362, 2023.

\bibitem{maddigan2023chat2vis}
Paula Maddigan and Teo Susnjak.
\newblock Chat2vis: Generating data visualisations via natural language using
  chatgpt, codex and gpt-3 large language models.
\newblock {\em IEEE Access}, 2023.

\bibitem{massarani2022role}
Luisa Massarani, Rosicler Neves, Graziele Scalfi, Antero Vin{\'\i}cius
  Portela~Firmino Pinto, Carla Almeida, Luis Amorim, Marina Ramalho, Luiz
  Bento, Monica Santos~Dahmouche, Renata Fontanetto, et~al.
\newblock The role of mediators in science museums: An analysis of
  conversations and interactions of brazilian families in free and mediated
  visits to an interactive exhibition on biodiversity.
\newblock {\em International Journal of Research in Education and Science},
  8(2):328--361, 2022.

\bibitem{marion}
Peter Mindek, David Kou{\v{r}}il, Johannes Sorger, Daniel Toloudis, Blair
  Lyons, Graham Johnson, M~Eduard Gr{\"o}ller, and Ivan Viola.
\newblock Visualization multi-pipeline for communicating biology.
\newblock {\em IEEE Transactions on Visualization and Computer Graphics},
  24(1):883--892, 2017.

\bibitem{FacilitatingConvInteraction}
Rishab Mitra, Arpit Narechania, Alex Endert, and John Stasko.
\newblock Facilitating conversational interaction in natural language
  interfaces for visualization.
\newblock In {\em 2022 IEEE Visualization and Visual Analytics (VIS)}, pages
  6--10, 2022.

\bibitem{narechania2020nl4dv}
Arpit Narechania, Arjun Srinivasan, and John Stasko.
\newblock Nl4dv: A toolkit for generating analytic specifications for data
  visualization from natural language queries.
\newblock {\em IEEE Transactions on Visualization and Computer Graphics},
  27(2):369--379, 2020.

\bibitem{nguyen2021modeling}
Ngan Nguyen, Ond{\v{r}}ej Strnad, Tobias Klein, Deng Luo, Ruwayda Alharbi,
  Peter Wonka, Martina Maritan, Peter Mindek, Ludovic Autin, David~S Goodsell,
  et~al.
\newblock Modeling in the time of covid-19: Statistical and rule-based
  mesoscale models.
\newblock {\em IEEE transactions on visualization and computer graphics},
  27(2):722, 2021.

\bibitem{openai_chatgpt_2022}
OpenAI.
\newblock Openai: Introducing chatgpt.
\newblock \url{https://openai.com/blog/chatgpt}, 2022.
\newblock Accessed: March 27, 2023.

\bibitem{openai2023gpt4}
OpenAI.
\newblock Gpt-4 technical report, 2023.

\bibitem{pooryousef2023working}
Vahid Pooryousef, Maxime Cordeil, Lonni Besan\c{c}on, Christophe Hurter, Tim
  Dwyer, and Richard Bassed.
\newblock Working with forensic practitioners to understand the opportunities
  and challenges for mixed-reality digital autopsy.
\newblock In {\em Proc.\ CHI}, CHI '23, New York, NY, USA, 2023. Association
  for Computing Machinery.

\bibitem{reinschluessel2022virtual}
Anke~V Reinschluessel, Thomas Muender, Daniela Salzmann, Tanja Doering, Rainer
  Malaka, and Dirk Weyhe.
\newblock Virtual reality for surgical planning--evaluation based on two liver
  tumor resections.
\newblock {\em Frontiers in Surgery}, 9:821060, 2022.

\bibitem{Rheingans2020}
Penny Rheingans, Helen-Nicole Kostis, Paulo~A. Oemig, Geraldine~B. Robbins, and
  Anders Ynnerman.
\newblock {\em Reaching Broad Audiences in an Educational Setting}, pages
  365--380.
\newblock Springer International Publishing, Cham, 2020.

\bibitem{scanlon2011technology}
Eileen Scanlon, Stamatina Anastopoulou, Lucinda Kerawalla, and Paul Mulholland.
\newblock How technology resources can be used to represent personal inquiry
  and support students' understanding of it across contexts.
\newblock {\em Journal of Computer Assisted Learning}, 27(6):516--529, 2011.

\bibitem{schiele2021science}
Bernard Schiele.
\newblock Science museums and centres: evolution and contemporary trends.
\newblock In {\em Routledge handbook of public communication of science and
  technology}, pages 53--76. Routledge, 2021.

\bibitem{schonborn2010bridging}
Konrad~J Sch{\"o}nborn and Trevor~R Anderson.
\newblock Bridging the educational research-teaching practice gap: Foundations
  for assessing and developing biochemistry students' visual literacy.
\newblock {\em Biochemistry and molecular biology education}, 38(5):347--354,
  2010.

\bibitem{seale2023education}
Nellie Seale.
\newblock Education, entertainment, and engagement in museums in the digital
  age.
\newblock In {\em Companion Proceedings of the Annual Symposium on
  Computer-Human Interaction in Play}, pages 326--329, 2023.

\bibitem{shaby2019engagement}
Neta Shaby, Orit Ben-Zvi~Assaraf, and Tali Tal.
\newblock Engagement in a science museum--the role of social interactions.
\newblock {\em Visitor Studies}, 22(1):1--20, 2019.

\bibitem{shaby2019examination}
Neta Shaby, Orit Ben-Zvi~Assaraf, and Tali Tal.
\newblock An examination of the interactions between museum educators and
  students on a school visit to science museum.
\newblock {\em Journal of Research in Science Teaching}, 56(2):211--239, 2019.

\bibitem{shahab2023virtual}
Hamza Shahab, Mozard Mohtar, Ezlika Ghazali, Philipp~A Rauschnabel, and Andrea
  Geipel.
\newblock Virtual reality in museums: does it promote visitor enjoyment and
  learning?
\newblock {\em International Journal of Human--Computer Interaction},
  39(18):3586--3603, 2023.

\bibitem{Sobel:2021:RBPC}
David~M Sobel, Susan~M Letourneau, Cristine~H Legare, and Maureen Callanan.
\newblock Relations between parent--child interaction and children’s
  engagement and learning at a museum exhibit about electric circuits.
\newblock {\em Developmental Science}, 24(3):e13057, 2021.

\bibitem{InChorus}
Arjun Srinivasan, Bongshin Lee, Nathalie Henry~Riche, Steven~M. Drucker, and
  Ken Hinckley.
\newblock Inchorus: Designing consistent multimodal interactions for data
  visualization on tablet devices.
\newblock In {\em Proc.\ CHI}, CHI '20, page 1–13, New York, NY, USA, 2020.
  Association for Computing Machinery.

\bibitem{alpaca}
Rohan Taori, Ishaan Gulrajani, Tianyi Zhang, Yann Dubois, Xuechen Li, Carlos
  Guestrin, Percy Liang, and Tatsunori~B. Hashimoto.
\newblock Stanford alpaca: An instruction-following llama model.
\newblock \url{https://github.com/tatsu-lab/stanford_alpaca}, 2023.

\bibitem{alpaca-lora}
Tloen.
\newblock Alpaca lora library.
\newblock \url{https://github.com/tloen/alpaca-lora}, 2023.
\newblock 08-August-2023.

\bibitem{Tsang:2002:BOOM}
Michael Tsang, George~W Fitzmaurice, Gordon Kurtenbach, Azam Khan, and Bill
  Buxton.
\newblock Boom chameleon: Simultaneous capture of {3D} viewpoint, voice and
  gesture annotations on a spatially-aware display.
\newblock In {\em Proc.\ UIST}, pages 111--120, New York, 2002. ACM.

\bibitem{Llama-2-70b-instruct-v2}
Upstage.
\newblock Llama-2-70b-instruct-v2.
\newblock \url{https://huggingface.co/upstage/Llama-2-70b-instruct-v2}, 2023.
\newblock 10-August-2023.

\bibitem{LLaMa-30b-instruct}
Upstage.
\newblock Llama-30b-instruct.
\newblock \url{https://huggingface.co/upstage/llama-30b-instruct}, 2023.
\newblock 10-August-2023.

\bibitem{wang2022towards}
Yun Wang, Zhitao Hou, Leixian Shen, Tongshuang Wu, Jiaqi Wang, He~Huang,
  Haidong Zhang, and Dongmei Zhang.
\newblock Towards natural language-based visualization authoring.
\newblock {\em IEEE Transactions on Visualization and Computer Graphics},
  29(1):1222--1232, 2022.

\bibitem{NanovisT4Model}
John Winfer, Aeliya Syed, Leon~Thistle Paul~Ekers, Ngan Nguyen, Ondrej Strnad,
  David Goodsell, Ivan Viola, and Deng Luo.
\newblock T4 model.
\newblock \url{https://www.nanovis.org/T4-model.html}.
\newblock (Accessed on 09/07/2021).

\bibitem{Xu:2022}
Ningning Xu, Yue Li, Jie Lin, Lingyun Yu, and Hai-Ning Liang.
\newblock User retention of mobile augmented reality for cultural heritage
  learning.
\newblock In {\em 2022 IEEE International Symposium on Mixed and Augmented
  Reality Adjunct (ISMAR-Adjunct)}, pages 447--452, 2022.

\bibitem{yao2020molecular}
Hangping Yao, Yutong Song, Yong Chen, Nanping Wu, Jialu Xu, Chujie Sun, Jiaxing
  Zhang, Tianhao Weng, Zheyuan Zhang, Zhigang Wu, et~al.
\newblock Molecular architecture of the sars-cov-2 virus.
\newblock {\em Cell}, 183(3):730--738, 2020.

\bibitem{yap2014structure}
Moh~Lan Yap and Michael~G Rossmann.
\newblock Structure and function of bacteriophage t4.
\newblock {\em Future microbiology}, 9(12):1319--1327, 2014.

\bibitem{Ynnerman2020}
Anders Ynnerman, Patric Ljung, and Alexander Bock.
\newblock {\em Reaching Broad Audiences from a Science Center or Museum
  Setting}, pages 341--364.
\newblock Springer International Publishing, Cham, 2020.

\bibitem{ynnerman:18:exploranation}
Anders Ynnerman, Jonas L{\"o}wgren, and Lena Tibell.
\newblock Exploranation: A new science communication paradigm.
\newblock {\em IEEE computer graphics and applications}, 38(3):13--20, 2018.

\bibitem{ynnerman:16:inside}
Anders Ynnerman, Thomas Rydell, Daniel Antoine, David Hughes, Anders Persson,
  and Patric Ljung.
\newblock Interactive visualization of {3D} scanned mummies at public venues.
\newblock {\em Commun. ACM}, 59(12):72--81, December 2016.

\end{thebibliography}

\end{document}